\documentclass[journal]{IEEEtran}
\usepackage[utf8]{inputenc}
\usepackage{amsmath, amssymb}
\DeclareMathOperator*{\argmin}{arg\,min}
\DeclareMathOperator*{\argmax}{arg\,max}
\usepackage{cite}
\usepackage{graphicx}
\usepackage{subfigure}
\usepackage{url}
\usepackage{xcolor}
\usepackage{siunitx}
\usepackage{gensymb}
\usepackage{textcomp}
\usepackage{algorithm}
\usepackage[numbers,sort&compress]{natbib}
\usepackage{amsmath}
\usepackage{bm}
\usepackage{multirow,tabularx}
\usepackage{threeparttable}
\usepackage{makecell}
\usepackage{array}
\newcolumntype{P}[1]{>{\centering\arraybackslash}p{#1}}
\newcolumntype{M}[1]{>{\centering\arraybackslash}m{#1}}
\usepackage{color}
\usepackage{float}

\begin{document}

\title{MADRL-Based Rate Adaptation for 360$\degree$ Video Streaming with Multi-Viewpoint Prediction}

\author{Haopeng Wang,
        Zijian Long,
        Haiwei Dong,~\IEEEmembership{Senior Member,~IEEE},
        Abdulmotaleb El Saddik,~\IEEEmembership{Fellow,~IEEE}
\thanks{Manuscript received March 14, 2024; accepted May 3, 2024. (Corresponding author: Haiwei Dong.)}
\thanks{Haopeng Wang and Zijian Long are with the University of Ottawa, Ottawa, ON, K1N 6N5, Canada (e-mail: \{hwang266, zlong038\}@uottawa.ca).}
\thanks{Haiwei Dong is with Huawei Canada, Ottawa, ON, K2K 3J1 and the University of Ottawa, Ottawa, ON, K1N 6N5, Canada (e-mail: haiwei.dong@ieee.org).}
\thanks{Abdulmotaleb El Saddik is with the University of Ottawa, Ottawa, ON, K1N 6N5, Canada and MBZUAI, Abu Dhabi, UAE (email: elsaddik@uottawa.ca).}}

\markboth{IEEE Internet of Things Journal}%
{Shell \MakeLowercase{\textit{et al.}}: A Sample Article Using IEEEtran.cls for IEEE Journals}

\maketitle
\begin{abstract}
Over the last few years, 360$\degree$ video traffic on the network has grown significantly. A key challenge of 360$\degree$ video playback is ensuring a high quality of experience (QoE) with limited network bandwidth. Currently, most studies focus on tile-based adaptive bitrate (ABR) streaming based on single viewport prediction to reduce bandwidth consumption. However, the performance of models for single-viewpoint prediction is severely limited by the inherent uncertainty in head movement, which can not cope with the sudden movement of users very well. This paper first presents a multimodal spatial-temporal attention transformer to generate multiple viewpoint trajectories with their probabilities given a historical trajectory. The proposed method models viewpoint prediction as a classification problem and uses attention mechanisms to capture the spatial and temporal characteristics of input video frames and viewpoint trajectories for multi-viewpoint prediction.
After that, a multi-agent deep reinforcement learning (MADRL)-based ABR algorithm utilizing multi-viewpoint prediction for 360$\degree$ video streaming is proposed for maximizing different QoE objectives under various network conditions. We formulate the ABR problem as a decentralized partially observable Markov decision process (Dec-POMDP) problem and present a MAPPO algorithm based on centralized training and decentralized execution (CTDE) framework to solve the problem. The experimental results show that our proposed method improves the defined QoE metric by up to 85.5\% compared to existing ABR methods. 

\end{abstract}

\section{Introduction}
Virtual Reality (VR) technology has reached a new level of maturity after years of development. As one of the key applications of VR, {360\degree} video is becoming increasingly popular \cite{9560167}. 
In 360$\degree$ video playback, users utilize mobile devices or HMDs (Head-mounted displays) to watch different contents by adjusting their viewpoint. The 360$\degree$ video player calculates the viewpoint position using different sensors, such as cameras and gyroscopes in the devices and displays the viewing area accordingly \cite{10.1145/2980055.2980056}. 360$\degree$ video has been used widely in football matches, museums, and concerts to attract audience attention \cite{9024132}. In the first 1.5 years after 360$\degree$ video support was launched on Facebook, more than 1 million 360$\degree$ videos were posted. A study found that 360$\degree$ videos increase web clicks by 8 times, and the viewing time of 360$\degree$ videos is 29\% longer than that of traditional videos \cite{8779683}.

\begin{figure*}[htbp]
\centering
\includegraphics[width=0.95\textwidth]{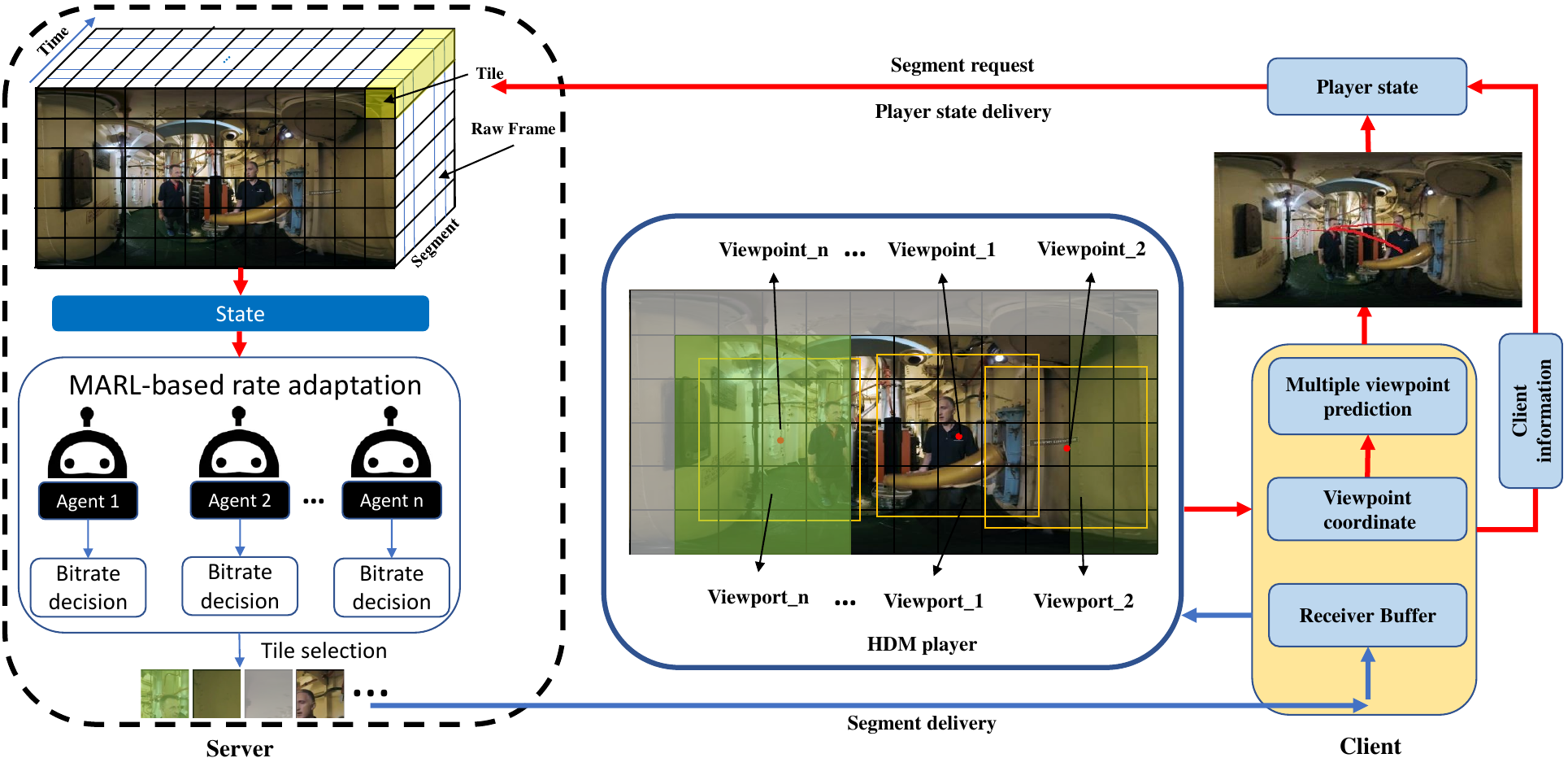}
\caption{The system architecture of the proposed MADRL-based 360$\degree$ video streaming system using predicted multiple viewpoint trajectories. An agent determines a bitrate for a viewport region. A tile's quality varies depending on the region's viewing probability and client information when it is streamed to a client. }
\label{view}
\end{figure*}

As 360$\degree$ video requires high bandwidth, streaming 360$\degree$ videos with high quality of experience (QoE) is quite challenging. Additionally, network conditions usually change during playback, which makes maintaining the quality difficult. To reduce bandwidth requirements and ensure a high QoE during video playback, many tile-based adaptive bitrate streaming approaches have been proposed \cite{9024132,10.1145/3394171.3413751, 7823595, 10.1145/3123266.3123291,hosseini2017view, 9854081, 9514569}. A 360$\degree$ video can be divided into several segments temporally and tiles spatially. A user's viewpoint trajectory can be predicted based on the historical trajectories of other users, as trajectories from other users present similar spatial and temporal characteristics. When a video is streamed to a client, the system determines the future viewpoint trajectory by using a variety of techniques, such as LR (linear regression) \cite{10.1145/2980055.2980056}, clustering algorithm \cite{8613652}, CNN (convolutional neural network) \cite{10.1145/3240508.3240669}, LSTM (long short-term memory)  \cite{9395242}, and DRL (deep reinforcement learning) \cite{8418756}. With information on predicted viewpoints, the system prioritizes tiles based on specific benchmarks, such as distance to the viewpoint. The bitrate determination for each tile is formulated as a QoE optimization problem, which is an NP-hard problem \cite{9234071}. Therefore, many methods based on the heuristic algorithm are proposed, such as beam search \cite{9024132}, dynamic programming \cite{10.1145/3240508.3240556} and greedy algorithm \cite{10.1145/3123266.3123291}. However, the heuristic solutions are time-consuming, and hard to achieve optimal solutions across various network conditions. Hence, DRL-based methods are expected to address the above challenges due to their strong adaptability to complex environments. The DRL method observes the environment (e.g., network conditions, buffer status, and video playback information) and takes actions to optimize QoE. The DRL agent receives the reward for the previous decision, guiding it to make better decisions over time. After learning from past experience by trial-and-error, it gradually converges to the optimal policy. The DRL-based ABR streaming significantly enhances video delivery by dynamically adapting to fluctuating network conditions and optimizing the viewing experience, providing an adaptive, scalable, and efficient solution for superior video quality.

However, the existing methods only predict a single viewpoint trajectory using historical trajectories or video content, which has challenges in handling sudden movements. On the other hand, a 360$\degree$ video is usually split into dozens of tiles, which makes the DRL difficult to apply in tile-based 360$\degree$ video streaming systems. For a 360$\degree$ video with $M$ tiles and $N$ bitrate levels, if the SADRL-based method determines the bitrate for all tiles simultaneously, it has an action space of $N^M$ dimension. If the MADRL (multi-agent reinforcement learning)-based method has an agent for each tile, which would be $M$ agent in the system. For example, if a 360$\degree$ video has 72 tiles and 6 bitrate levels, there would be a $6^{72}$ dimension action space for the SADRL-based method and a 6 dimension action space but 72 agents for the MADRL-based method.

Since network conditions vary and users move their heads unexpectedly without following the predicted trajectory, it is challenging for the tile-based ABR streaming technology to give users a high QoE. In the SADRL, a single agent is tasked with adjusting the bitrates for all tiles. However, as the number of tiles along with the state and action spaces increases substantially, SADRL faces significant challenges in learning efficiently and making the best decisions for the dynamic changes of user behavior and network conditions. Therefore, to deal with the above issues, we propose a multi-viewpoint prediction strategy based on a multimodal spatial-temporal attention transformer to handle users' sudden head movements and a MADRL-based ABR streaming system based on multi-viewpoint prediction for fluctuating network conditions. The multi-trajectory prediction enhances the accuracy and reliability by not depending on a single, potentially incorrect path. Consequently, the MADRL approach with the CTDE framework based on the multi-viewpoint prediction is more capable of scaling to manage a high volume of tiles efficiently. Within MADRL, agents have the flexibility and responsiveness to adjust their strategies to cope with complex user behaviors and fluctuating network conditions by effectively managing the interdependencies between tiles. To reduce the number of MADRL agents, the following rules are applied to the proposed system to determine bitrates for the tiles. 
\begin{itemize}
    \item Rule 1: The 360$\degree$ video is divided into different viewport regions according to viewpoint positions, as viewports could be overlapped if the distance between viewpoints is too short. The overlapped area is assigned to the viewport region with a higher viewing probability. 
    \item Rule 2: Since the area outside the viewport regions has the lowest viewing probability, tiles outside the viewport regions are assigned the lowest bitrate.
    \item Rule 3: All tiles in the same viewport region are assigned the same bitrate, so the number of agents in the MADRL is reduced significantly and determined by the number of viewport regions.
\end{itemize}

The proposed ABR streaming system for 360$\degree$ video is shown in Fig. \ref{view}. A 360$\degree$ video is split into segments temporally and tiles spatially and stored in a server. A user's viewpoint is initialized in the center of the 360$\degree$ video when the user begins viewing the 360$\degree$ video. With the 360$\degree$ video playback, the viewpoint prediction module predicts multiple future viewpoint trajectories with their viewing probabilities using historical viewing traces. The 360$\degree$ video is divided into three viewport regions based on the prediction results and a region containing all the rest tiles. The prediction results together with other client information (agent observation described in section IV) are fed into the MADRL-based rate adaptation module. By utilizing the information of the environment state, the rate adaptation module makes decisions on the bitrate for each viewport region. The tiles with different bitrates are then streamed to the client. The tiles in the grey region shown in Fig.\ref{view} are allocated with the lowest bitrate. Three viewport regions are allocated with various bitrates represented by different transparencies. Noticeably, our approach predicts multiple viewpoints along with their respective probabilities and assigns a uniform bitrate to a viewport area. However, we observe that this could lead to bitrate conflicts for the intersections of different viewport areas. To address this issue, we adopt a policy where the bitrate of an intersection is determined according to the viewpoint with the highest probability. As illustrated in Fig. \ref{view}, since viewpoint 1 has a higher probability than viewpoint 2, the intersection's bitrate is determined by the agent of viewport 1 utilizing the probability of viewpoint 1. Following is a summary of the contribution of this paper:
\begin{itemize}
    \item A new 360$\degree$ video ABR streaming strategy is proposed, in which multi-viewpoint prediction based on attention mechanism and adaptive bitrate streaming based on MADRL are jointly employed to improve the performance of the 360$\degree$ video streaming system.
    \item A multimodal spatial-temporal attention transformer is proposed for multi-viewpoint prediction. The model utilizes spatial and temporal attention mechanisms to capture information about input video frames and viewpoint trajectories. The obtained representations are fused to predict multiple viewpoint trajectories. 
    \item The adaptive bitrate streaming is formulated as a decentralized partially observable Markov decision process (Dec-POMDP) optimization problem maximizing a defined QoE metric. A MADRL method with the MAPPO (multi-agent proximal policy optimization) algorithm based on the centralized training and decentralized execution framework (CTDE) is proposed to obtain the optimal Dec-POMDP solution. 
    \item A comparison between the proposed method and other existing algorithms for 360$\degree$ video streaming is conducted. The experimental results show that the proposed method achieves better performance for the various metrics under different network conditions.
\end{itemize}

\section{Related Work}
\subsection{Viewpoint Prediction}
Many tile-based ABR allocation strategies utilizing the future viewpoint information have been proposed to improve the performance of 360$\degree$ video streaming systems. Viewpoint prediction is crucial for reducing bandwidth costs and providing high quality. A user's viewpoints can be predicted by utilizing LR with the user's historical viewpoints \cite{10.1145/2980055.2980056, 7823595}. Meanwhile, some probabilistic models are proposed to estimate the distribution of prediction error to improve the performance of the LR method \cite{10.1145/3123266.3123291,8351404}. However, LR-based methods assume the linear movement of the head, which is a strong assumption and causes high bias. Therefore, many methods are proposed to extract temporal and spatial features from different users' viewpoint trajectories, which achieves better performance and dominates existing 360$\degree$ video streaming systems. Since viewpoint trajectories of a video from various users present similar spatial and temporal characteristics, the current user's viewpoint trajectory can be predicted based on historical data from other users. Petrangeli et al. \cite{8613652} cluster similar trajectories together using the spectral clustering algorithm. A different function is computed for each cluster and predicts future viewpoint positions. Similarly, Taghavi et al. \cite{10.1145/3386290.3396934} cluster the viewpoint trajectories from previous users into different groups. Based on the extrapolation of the quaternions, the user' trajectory is matched to one of the clusters. Viewpoints are predicted using the cluster center.

Additionally, many deep learning-based methods are proposed to predict viewpoints. Zhang et al. \cite{9024132} build three LSTM models and use the mean of prediction results as the final prediction. Fu et al. \cite{9234071} adopt LSTM and self-attention mechanism to predict viewpoint. Chao et al. \cite{9733647} utilize the transformer encoder to predict viewpoints. To enhance the accuracy of viewpoint prediction, viewpoint trajectory combining more information, such as video content and saliency map, are fed into a deep learning model. Xu et al. \cite{8418756} proposed a DRL-based method to predict head movement. The method takes the 360$\degree$ video content and past viewport trajectory as input and optimizes the difference of actions between the agent and user through the agent's actions. Nguyen et al. \cite{10.1145/3240508.3240669} proposed a CNN model to predict saliency maps and an LSTM model to predict future viewpoints with the predicted saliency maps and head orientation map. Romero et al. \cite{9395242} designed an LSTM model that utilizes previous viewpoint positions and saliency maps to predict future viewpoints.

All the aforementioned methods predict a single-viewpoint trajectory. The only work we found for multiple-viewpoint prediction is proposed by Guimard et al. \cite{10.1145/3524273.3528176}. By analyzing public viewpoint data, they demonstrate the need for multiple-viewpoint prediction due to various possible futures for similar past trajectories. A discrete variational learning method is proposed for multiple-viewpoint predictions. However, their method only generates multiple trajectories without probabilities while an extra likelihood estimator is built to obtain the probabilities of the prediction results. In this paper, we propose a transformer-based method to predict multiple-viewpoint trajectories with their viewing probabilities by treating viewpoint prediction as a classification problem. The probabilities indicate the importance of the predicted trajectories, which are used to determine the bitrate in our system.
\subsection{Rate Adaptation}
The ABR streaming can be formulated as a QoE optimization problem and solved by various heuristic algorithms. As heuristic algorithms are time-intensive and struggle to find the best solutions across different network conditions. Many SADRL-based methods are proposed to address this challenge.
Due to the dimension of action space increasing with the tiles and bitrate levels, an ABR strategy that sequentially determines bitrate for tiles one by one is presented, where A3C \cite{DBLP:journals/corr/MnihBMGLHSK16} algorithm is used \cite{9234071,9226435,9423318}. To further reduce the complexity of action space, bitrate adaptation based on the viewport region is proposed. Zhang et al. \cite{8737361} split the 360$\degree$ video into two regions: viewport and rest. The tiles inside the viewport have the same bitrate which is determined by a SADRL model with A3C, while the rest tiles have the lowest bitrate. Similarly, Kan et al. \cite{9419061} split the 360$\degree$ video into three regions: viewport, marginal and invisible region, and present a SADRL model with the A3C algorithm to determine the bitrates for three regions at the same time. Wei et al. \cite{9351629} propose a two-step strategy to determine the bitrate of tiles. A SADRL model first determines the bitrate of the segment. The bitrates of tiles are determined using game theory considering the view prediction and segment bitrate. Feng et al. \cite{9838819} classify the tiles inside the viewport into different levels and utilize the PPO algorithm to determine the bitrate for the tiles inside the viewport. However, all existing methods are based on SADRL, which usually obtains local optimal for the bitrate determination without globally considering the existence of other tiles. In this paper, we formulate the 360$\degree$ video streaming as a Dec-POMDP optimization problem and propose a MADRL method with MAPPO algorithm to globally determine the bitrate for tiles based on the multi-viewpoint prediction from the transformer method. 

\section{Viewpoint Prediction}
\begin{figure*}[htbp]
\centering
\includegraphics[width=0.95\textwidth]{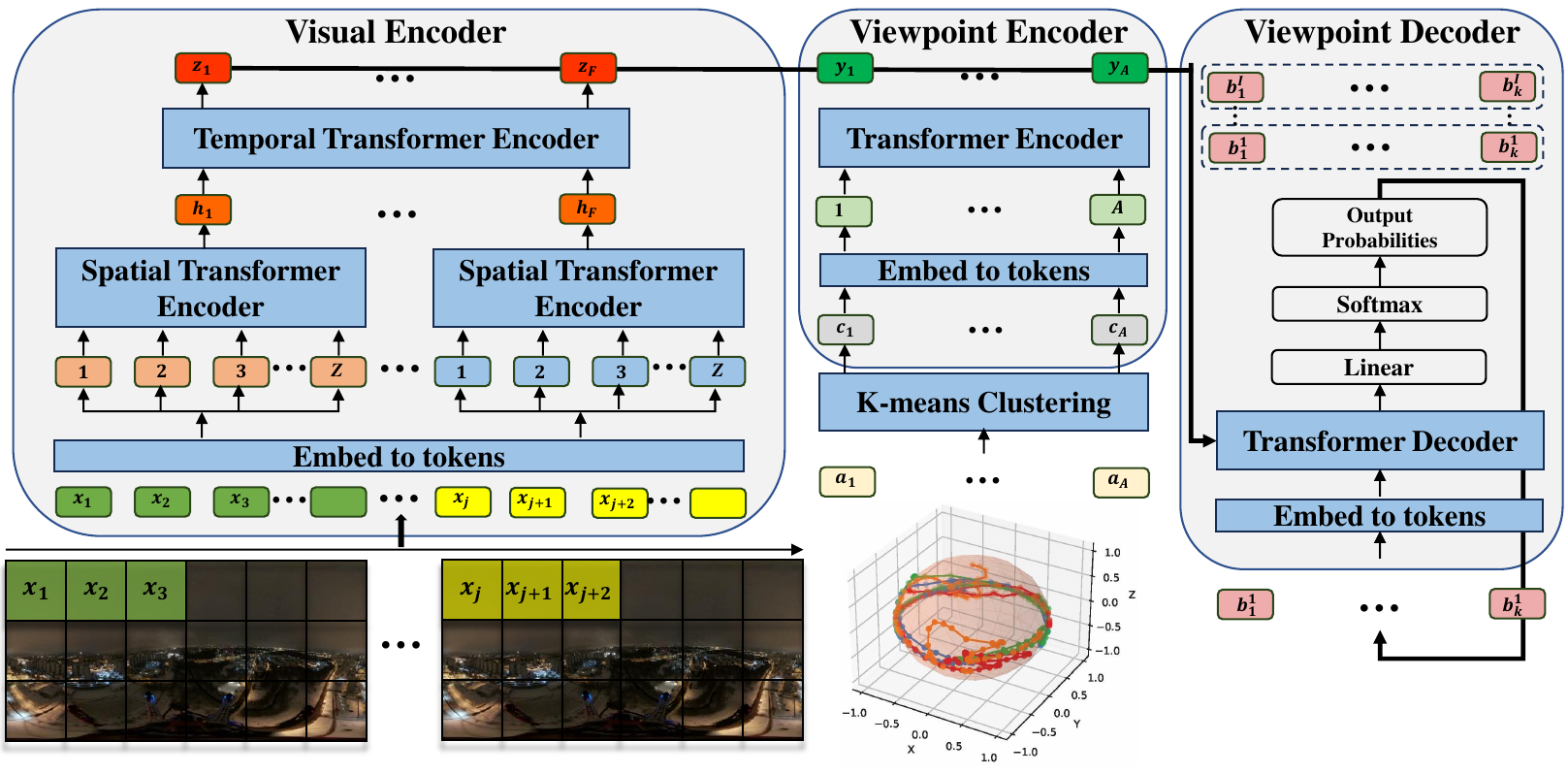}
\caption{The architecture of the multimodal spatial-temporal attention transformer model for multi-viewpoint prediction with classification method.}
\label{transformer}
\end{figure*}
In this paper, a multimodal spatial-temporal attention transformer is proposed to predict multiple viewpoint trajectories. Given a historical viewpoint trajectory, $\{a_m\}^A_{m=1}$, where position $a_m \in \mathbb{R}^3$, and a sequence of video frames $\{f_j\}^F_{j=1}$, where frame $f_j \in \mathbb{R}^{H \times W \times C}$, and $H$, $W$ and $C$ indicate the height, width, and channel number. The goal is to predict a user's $I$ viewpoint trajectories $\{b^i_r\}^{B}_{r=1}$, where $i$ represents $i$th predicted trajectory, $i = 1, 2, ..., I$. As shown in Fig. \ref{transformer}, the model consists of a visual encoder, a viewpoint encoder, and a viewpoint decoder. The encoder-decoder architecture proposed by Vaswani et al. \cite{10.5555/3295222.3295349} is utilized in our model. Each module is described next.

\subsection{Visual Encoder}
Every frame is divided into $Z$ patches, and patch $x_i \in \mathbb{R}^{h \times w}$, $i=1,...,Z$, where $h$ and $w$ are height and width, respectively. The patch embedding and positional embedding are used to embed patches into tokens, as with the ViT \cite{vit}. The spatial encoder only models interactions between tokens extracted from the same frame. The spatial attention score of patch $x_i$ of frame $f_j$ is given by: 
\begin{equation}
    Attention_{spatial}=softmax(\frac{q_{(x_i,f_i)}}{\sqrt{d_k}} \cdot \{k_{(x_\alpha,f_j)}\}^T_{\alpha=1,...,Z})
\end{equation}
where $q_{(x_i,f_j)}$ and $k_{(x_\alpha,f_j)}$ represent query vector of patch $x_i$ and key vector of patch $x_\alpha$ in frame $f_j$. The $d_k$ indicates the embedding dimension.
Representations $\{h_i\}^F_{i=1}$,  $h_i \in \mathbb{R}^{d_k}$, of temporal frames are then obtained. The frame-level representations are concatenated into, $H \in \mathbb{R}^{F \times {d_k}}$, and then processed by a temporal encoder to model interactions between tokens from different temporal frames. The temporal attention score of frame $f_j$ is given by:
\begin{equation}
    Attention_{temporal}=softmax(\frac{q_{(f_j)}}{\sqrt{d_k}} \cdot \{k_{(f_\beta)}\}^T_{\beta=1,...,F})
\end{equation}
where $q_{(f_j)}$ and $k_{(f_\beta)}$ are the query vector of frame $f_j$ and key vector of frame $f_\beta$. 
The output tokens $\{z_j\}_{j=1}^{F}$, $z_j \in \mathbb{R}^{d_k}$ are then obtained.  
\subsection{Viewpoint Encoder}
The trajectory prediction is treated as a classification problem, rather than a regression problem that directly predicts coordinates. The K-means clustering algorithm classifies the original viewpoints $\{a_m\}^A_{m=1}$ into different groups to get centroids. A centroid $c_i$ is then embedded into a token with word embedding and position embedding. The viewpoint encoder models the relationship between tokens extracted from historical viewpoints. The output tokens $\{y_m\}_{m=1}^A$, $y_m \in \mathbb{R}^{d_k}$ are obtained after viewpoint encoder.
\subsection{Viewpoint Decoder}
The viewpoint decoder generates the viewpoint set $\{b_r\}^{B}_{r=1}$ with the encoder tokens that are obtained by concatenating the visual tokens $\{z_j\}_{j=1}^{F}$  and viewpoint tokens $\{y_m\}_{m=1}^A$. At each autoregressive step $k$, the viewpoint decoder relies on a causal transformer decoder that cross-attends with the encoder outputs and self-attends with tokens generated in previous steps to generate a probability vector for all categories with the softmax activation function, where each element of the probability vector represents the predicted probability for a specific viewpoint position. The viewpoints with top-$I$ probabilities are sampled as the multi-viewpoint prediction.

\section{Transmission Rate Adaptation}
\subsection{QoE Metric Design}
We assume that a 360$\degree$ video is divided into $M$ tiles and $F$ segments with the same duration $\Delta t$. Here $\Delta t$ is set to $1$ second, $t$-th segment, $t \in \{1,2,...,M\}$, indicates the segment in time step $t$. {Since $I$ }viewpoint trajectories are predicted,  the 360$\degree$ video contains $I$ viewport regions and a rest region composed of the rest tiles outside viewport regions. For $t$-th segment, the viewing probability of viewport region $i$ is denoted as $\psi_{t}^{i}$. The buffer occupancy at time $t$ is denoted as $b_t$. The network throughput at time $t$ is defined as $\varphi_t$. The bitrate of the $m$-th tile, $m \in \{1,2,...,M\}$, is denoted as $r_{t}^m$. The size of the $m$-th tile with bitrate $r_{t}^m$ is denoted as $\Phi(tile_{t}^m)$. Therefore, the total size of segment $t$ is denoted as ${\Phi(Seg_t)}$, which can be computed by

\begin{equation}
    \Phi(Seg_t) = \sum_{m=1}^{M}\Phi(tile_{t}^m)
\end{equation}

The size of the region $i$ is denoted as $\Phi(Reg_t^{i})$ computed by
\begin{equation}
    \Phi(Reg_t^{i}) = \sum_{m \in {Reg_t^{i}}} \Phi(tile_{t}^m)
\end{equation}
where $Reg_t^{i}$ indicates all tiles of region $i$.

In this paper, the QoE model of the segment $t$ is defined by considering the following aspects:
\begin{itemize}
    \item Viewport quality: Since a user only views the content inside the viewport region, the quality of the viewport region is critical to the QoE. As we have $I$ viewport regions, the viewport quality is defined as the weighted average quality of viewport regions.
    \begin{equation}
        QoE_t^1 = \sum_{i=1}^{I} \psi_t^{i} \cdot q_t^{i}
    \end{equation}
    where $ q_t^j$ represents the quality of $i$-th viewport region of segment $t$. Here, the quality is defined as the bitrate, since users perceive higher video quality when a higher bitrate is selected.
    \item Viewport temporal variation: Due to the possibility of dizziness and headaches caused by sudden changes in rate between consecutive viewports, viewport temporal variation is also considered, which is the difference between the same viewport regions of two successive segments. The viewport temporal variation is defined as the average variation of all viewport regions. 
    \begin{equation}
        QoE_t^2 = \sum_{i=1}^{I} \psi_t^{i} \cdot ( q_t^{i}-q_{t-1}^{i})
    \end{equation}
        \item Viewport spatial variation: Users may view different viewport regions with the corresponding probability, and quality changes between viewport regions may lead to distorted edges. Therefore, the spatial variation between regions is also considered and is expressed by
    \begin{equation}
        QoE_t^3 = \frac{1}{2}\sum_{i=1}^{I} \sum_{{j} \in U_t^{i}} \psi_t^{i} \cdot \psi_t^{j} \cdot ( q_t^{i}-q_t^{j})
    \end{equation}
    where $U_t^{i}$ indicates the set of all viewport regions excluding $i$.
    \item Rebuffering: The client would suffer from rebuffering when the download time of a segment exceeds the buffer occupancy. We denote rebuffer time as
    \begin{equation}
        QoE_t^4 = max(\frac{\Phi(Seg_t)}{\varphi_t}-b_t+\Delta t, 0)
    \end{equation}

\end{itemize}
The QoE is defined as a weighted sum according to the aforementioned analysis.
\begin{equation}
   QoE_t = \alpha_1 QoE^1_t - \alpha_2 QoE^2_t - \alpha_3 QoE^3_t - \alpha_4 QoE^4_t
    \label{qoe}
\end{equation}
where $\alpha_*$ are weighting parameters for the factors and adjusted according to the user's preference.
\subsection{MADRL Problem Formulation}

The bitrate decision process for multiple viewport regions can be formulated as a Dec-POMDP \cite{oliehoek2016concise}. A Dec-POMDP is defined by a tuple $(\mathbb{D}, \mathbb{S}, \mathbb{A}, T, \mathbb{O}, O, R, \gamma)$. 
\begin{itemize}
    \item $\mathbb{D}=\{1,...,n\}$ is a set of $n$ agents.
    \item $\mathbb{S}$ is a set of states.
    \item  $\mathbb{A}=\{\mathbb{A}^{1},...,\mathbb{A}^{n}\}$ is a set of joint actions, where $\mathbb{A}^{i}$ is a set of actions available to agent $i \in \mathbb{D}$.
    \item  $P$ is the transition probability function that describes how a joint action affects the environment.
    \item  $\mathbb{O}=\{\mathbb{O}^{1},...,\mathbb{O}^{n}\}$ is a set of joint observations, where $\mathbb{O}^{i}$ is a set of observations available to agent $i$;
    \item  $O$ is the observation probability function.
    \item  $R$ is the reward function that describes the reward to the whole agents given their decisions.
    \item $\gamma \in (0,1] $ is a discount factor. 
\end{itemize}

At each time stage $t$, each agent takes an action $a_t^i \in \mathbb{A}^i$ according to their observation $o_t^i$, leading to a joint action $\bm{a_t}=\{a_t^{1},...,a_t^n\}$, which causes the current global state $\bm{s_t} \in \mathbb{S}$ of the environment to move to next state $\bm{s_{t+1}} \in \mathbb{S}$ with transition probability function $P(\bm{s_{t+1}}|\bm{s_t}, \bm{a_t})$. At the same time, each agent receives a new observation $o_{t+1}^i \in \mathbb{O}_i$ based on the observation probability function $O(o_{t+1}^i|\bm{s_{t+1}}, \bm{a_t})$ and a reward is generated for the whole team based on the reward function $R(\bm{s_t},\bm{a_t})$. The discount factor balances the immediate and future rewards. The objective of the Dec-PODMP is to find a tuple of policies, called a joint policy $\bm{\pi}$, to maximize the global discounted expected cumulative rewards. The agent, state, action, and reward for the proposed ABR strategy are defined below. 

\textbf{Agent Set}: Each viewport region has its corresponding agent, which is responsible for determining the bitrate for the region based on observation.

\textbf{State and Observation Set}: Agent $i$ at time step $t$ has its local observation which is defined as $o_t^i = (\Vec{\sigma^i_t}, \Vec{\varphi_t^i}, \Vec{\omega_t^i}, \psi^i_t,  c_t^i, l_t^i, b_t^i)$. $\Vec{\sigma^i_t}$ is the download time for the previous $k$ segments; $\Vec{\varphi_t^i}$ is the network throughput for the previous $k$ segments; $\Vec{\omega_t^i}$ is a vector containing all available sizes of the viewport region for the next segment; $\psi_t^i$ is the viewport probability of viewport region $i$;  $c_t^i$ is the number of remaining segments; $l_t^i$ is the last bitrate, and $b_t^i$ the current buffer level. The global state $\bm{s_t}=\{o^1_t,...,o^n_t\}$ is the concatenation of observations of all agents.

\textbf{Action Set}: The action $a_t^i$ of agent $i$ at time $t$ is to determine the bitrate of the viewport region $i$ for the next segment. Therefore, the joint action is the concatenation of all actions, $\bm{a_t}=(a_t^1, ..., a_t^n)$.

\textbf{Reward}: The global reward due to joint action is denoted as $r_t$, which is the QoE metric (Eq. \ref{qoe}). The local reward of agent $i$ at time $t$ can be derived from the global reward, as $\psi_t^i \cdot q_t^j+\psi_t^j \cdot ( q_t^i-q_{t-1}^i)+\sum_{{j} \in U_t^{i}} \psi_t^{i} \cdot \psi_t^{j} \cdot ( q_t^{i}-q_t^{j})+QoE_t^4$.

\subsection{Multi-Agent PPO Solution}
A MADRL algorithm named MAPPO, which extends PPO (proximal policy optimization) \cite{DBLP:journals/corr/SchulmanWDRK17}, is presented for optimizing bitrate adaptation streaming of 360$\degree$ video in the multi-agent environment. The local policy of agent $i$ is defined as $\pi^i_{\theta_i}$. The joint policy indicates the probability of selecting a joint action $\bm{a}$ under a global state $\bm{s}$ and is denoted as $\bm{\pi_\theta(a|s)}$. Since each agent makes an individual decision at each time step, the joint policy is defined as $\bm{\pi_\theta(a|s)}=\prod_{i \in \mathbb{D}}\pi^i_{\theta_i}(a^i|s^i)$.

The goal of multiple agents is to collaboratively find the optimal joint policy $\bm{\pi_{\theta}}$ that maximizes the global discounted accumulated expected reward. The global discounted accumulated reward, also called return, after joint action $\bm{a_t}$ at time $t$ is defined as:
\begin{equation}
\begin{split}
        G_t=r_{t+1}+\gamma r_{t+2}+\gamma^2 r_{t+3}+...
        =\sum_{k=0}^{\infty}\gamma^{k}r_{t+k+1}
\end{split}
\end{equation}
The global state value function under the policy $\bm{\pi_{\theta}}$ for global state $\bm{s}$ is defined as the expected return:
\begin{equation}
    V^{\bm{\pi_{\theta}}}(\bm{s})=E_{\bm{\pi_{\theta}}}[G_t|\bm{s_t}=\bm{s}]
\end{equation}
The global state-action value function under the policy $\bm{\pi_{\theta}}$ is defined as:
\begin{equation}
\begin{split}
    Q^{\bm{\pi_{\theta}}}(\bm{s},\bm{a})&=E_{\bm{\pi_{\theta}}}[G_t|\bm{s_t}=\bm{s}, \bm{a_t}=\bm{a}] 
\end{split}
\end{equation}
Therefore, the global state value function is rewritten as:
\begin{equation}
    V^{\bm{\pi_{\theta}}}(\bm{s})=\sum_{\bm{a} \in \mathbb{A}}Q^{\bm{\pi_{\theta}}}(\bm{s},\bm{a})\bm{\pi_{\theta}}(\bm{a}|\bm{s})
\end{equation}
The objective function is then defined as:
\begin{equation}
\begin{split}
       J(\bm{\theta}) &= \sum_{\bm{s} \in \mathbb{S}}d^{\bm{\pi_{\theta}}}(\bm{s})V^{\bm{\pi_{\theta}}}(\bm{s}) \\
       &=\sum_{\bm{s} \in \mathbb{S}}d^{\bm{\pi_{\theta}}}(\bm{s})\sum_{\bm{a} \in \mathbb{A}}Q^{\bm{\pi_{\theta}}}(\bm{s},\bm{a})\bm{\pi_{\theta}}(\bm{a}|\bm{s})
\end{split}
\end{equation}
where $d^{\bm{\pi_{\theta}}}(\bm{s})$ is the stationary distribution of Markov chain for $\bm{\pi_{\theta}}$. To maximize the objective function, the gradient of $J(\bm{\theta})$ with respect to $\bm{\theta}$ is computed by:

\begin{equation}
\begin{split}
    \nabla_{\bm{\theta}}J(\bm{\theta})&=\nabla_{\bm{\theta}}\sum_{\bm{s} \in \mathbb{S}}d^{\bm{\pi_{\theta}}}(\bm{s})\sum_{\bm{a} \in \mathbb{A}}Q^{\bm{\pi_{\theta}}}(\bm{s},\bm{a})\bm{\pi_{\theta}}(\bm{a}|\bm{s}) \\
    &\propto \sum_{\bm{s} \in \mathbb{S}}d^{\bm{\pi_{\theta}}}(\bm{s})\sum_{\bm{a} \in \mathbb{A}}Q^{\bm{\pi_{\theta}}}(\bm{s},\bm{a})\nabla_{\bm{\theta}}\bm{\pi_{\theta}}(\bm{a}|\bm{s}) \\
    &=\sum_{\bm{s} \in \mathbb{S}}d^{\bm{\pi_{\theta}}}(\bm{s})\sum_{\bm{a} \in \mathbb{A}}\bm{\pi_{\theta}}(\bm{a}|\bm{s})Q^{\bm{\pi_{\theta}}}(\bm{s},\bm{a})\frac{\nabla_{\bm{\theta}}\bm{\pi_{\theta}}(\bm{a}|\bm{s})}{\bm{\pi_{\theta}}(\bm{a}|\bm{s})} \\
    &=E_{\bm{\pi}}[Q^{\bm{\pi_{\theta}}}(\bm{s},\bm{a})\nabla_{\bm{\theta}}\log \bm{\pi_{\theta}}(\bm{a}|\bm{s})]
\end{split}
\end{equation}
where $E_{\bm{\pi}}$ refers to $E_{\bm{s} \sim d^{\bm{\pi_{\theta}}},\bm{a} \sim {\bm{\pi_\theta}}}$ as distributions of state and action follow the policy $\bm{\pi_{\theta}}$.
We are only concerned with determining how much better an action is compared to others on average, rather than describing its absolute quality. Therefore, we focus on determining the relative advantage of a particular action with the concept of an advantage function, which can be defined by
\begin{equation}
    A^{\bm{\pi_{\theta}}}(\bm{s},\bm{a})=Q^{\bm{\pi_{\theta}}}(\bm{s},\bm{a})-V^{\bm{\pi_{\theta}}}(\bm{s})
\end{equation}
So we can rewrite the gradient of $J(\bm{\theta})$ using the advantage function \cite{sutton2018reinforcement}
\begin{equation}
    \nabla_{\bm{\theta}}J(\bm{\theta})=E_{\bm{\pi}}[A^{\bm{\pi_{\theta}}}(\bm{s},\bm{a})\nabla_{\bm{\theta}}\log \bm{\pi_{\theta}}(\bm{a}|\bm{s})]
\end{equation}
Due to the independence of each $\theta^i$ and $\bm{\pi_\theta}(\bm{a}|\bm{s})=\prod_{i \in \mathbb{D}}\pi^i_{\theta_i}(a^i|s^i)$, the partial derivative of $J(\bm{\theta})$ with respect to local parameter $\theta^i$ is written as:
\begin{equation}
    \nabla_{\theta^i}J(\bm{\theta})=E_{\bm{\pi}}[A^{\bm{\pi_{\theta}}}(\bm{s},\bm{a})\nabla_{\theta^i}\log \pi^i_{\theta^i}(a^i|s^i)]
    \label{e1}
\end{equation}
Since the gradient of $J(\bm{\theta})$ at $\bm{\theta}$ can be obtained from all the partial derivatives $\nabla_{\theta^1}J(\bm{\theta}),\nabla_{\theta^2}J(\bm{\theta}),...,\nabla_{\theta^n}J(\bm{\theta})$. If the stochastic gradient ascent of every agent's $\theta^i$ keeps a sufficiently small displacement of $\bm{\theta}$, we can improve the joint policy globally. 
\begin{figure}[htbp]
\centering
\includegraphics[width=0.47\textwidth]{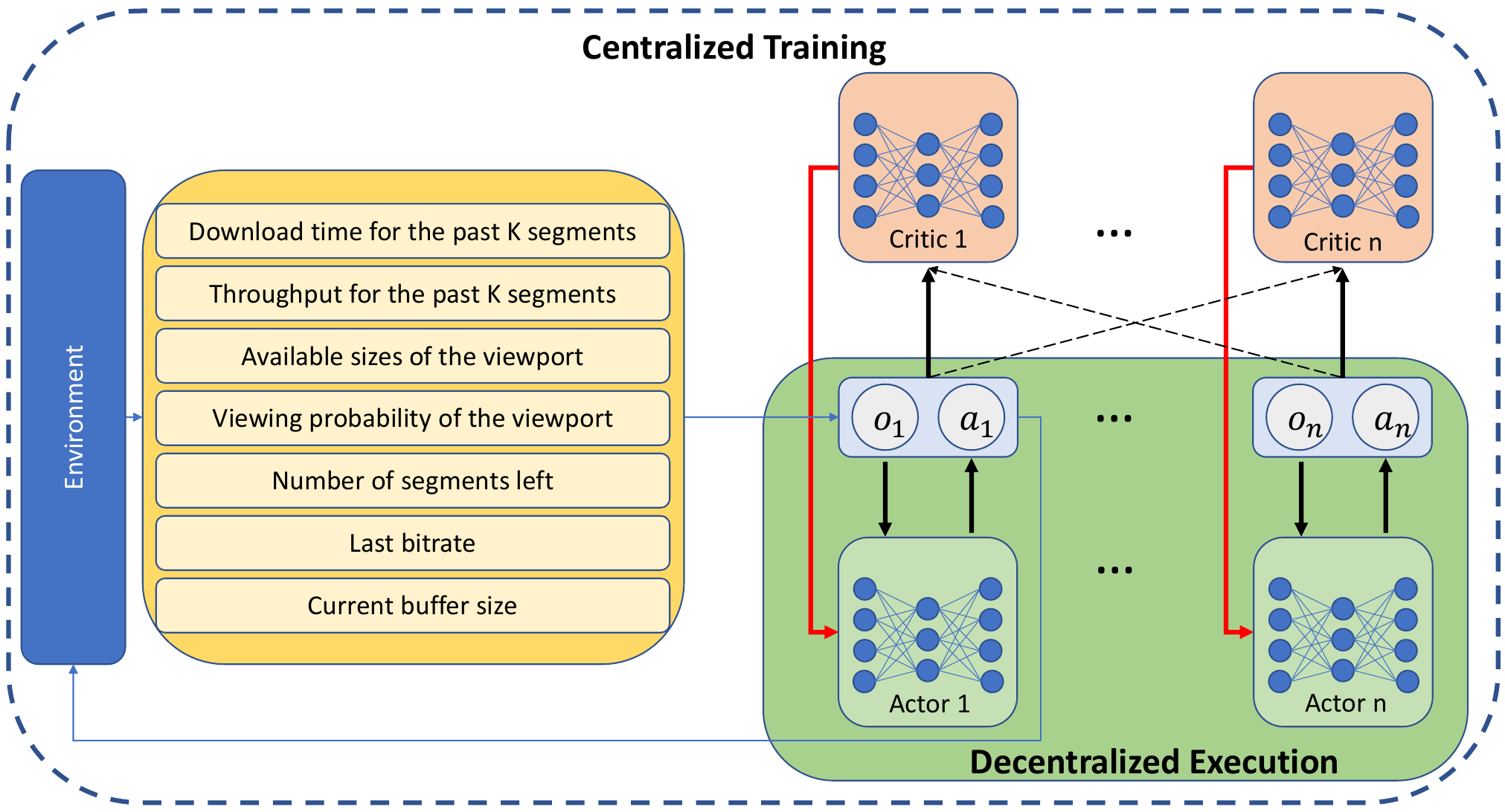}
\caption{The proposed MADRL with MAPPO algorithm based on CTDE framework.}
\label{ctde}
\end{figure}

Additionally, it can be seen that when the advantage function $A^{\bm{\pi_{\theta}}}(\bm{s},\bm{a})$ is given, $\nabla_{\theta^i}J(\bm{\theta})$ can be obtained locally, which allows us to apply the centralized training and distributed execution and extend the policy gradient algorithms in MADRL. As shown in Fig. \ref{ctde}, the actor of each agent is fed with its corresponding local observation containing various player information. The joint action and global observation are fed into critics. The centralized critics stabilize the training by considering the states and actions of all agents, which helps the actors make cooperative actions and avoid conflicts based on their partial information without additional information during the decentralized exploration and execution stages. In addition, an agent determines the bitrate for a viewport region with a viewing probability. The local and global rewards are weighted by viewport viewing probabilities. Different agents have different contributions to the global reward even with the same action. In this way, all agents are constrained by the viewing probability and reward, where they are encouraged to behave cooperatively and avoid conflicting behaviors.

The MAPPO is implemented according to Eq. \ref{e1}. The MAPPO maintains two separate networks for each agent: a policy network (actor) with parameter $\theta^i$ and a value function network (critic) with parameter $\phi^i$. When the probability ratio between old and new policies is clipped, the stochastic gradient ascent of each agent's $\theta^i$ can remain within a small range of $\bm{\theta}$ as much as possible.
During the training period, some trajectories can be collected from the environment. A trajectory with horizon $T$ can be defined as  $\tau=\{\bm{s}_0,\bm{a}_0,r_0,...,\bm{s}_{T-1},\bm{a}_{T-1},r_{T-1},\bm{s}_T\}$. At time step $t$, the estimated discounted return $\hat G_t$ on trajectory $\tau$ is calculated by 
\begin{equation}
    \hat G_t=\sum_{i=t}^{T-1}\gamma^{i-t}r_{t}+\gamma^{T-t}V_{\bm{\phi}}(\bm{s}_T)
\end{equation}
where $V_{\bm{\phi}}$ is the global value function with parameter $\bm{\phi}$.

The estimated global advantage function $\hat{A}(\bm{s}_t,\bm{a}_t)$ on trajectory $\tau$ can be calculated using the truncated generalized advantage estimation (GAE) \cite{DBLP:journals/corr/SchulmanWDRK17}:
\begin{equation}
\begin{split}
    \hat{A}(\bm{s}_t,\bm{a}_t)&=\delta_t+(\gamma \lambda)\delta_{t+1}+...+(\gamma \lambda)^{T-1-t}\delta_{T-1}\\
    &=\sum_{l=0}^{T-1-t}(\lambda\gamma)^l\delta_{t+l}
\end{split}
\end{equation}
where $\delta_t$ is the temporal difference residual and defined as $\delta_t=r_t+\gamma V_{\bm{\phi}}(\bm{s}_{t+1})-V_{\bm{\phi}}(\bm{s}_{t})$. The hyperparameter $\lambda \in [0,1]$ is used to balance the bias-variance trade-off. 

To improve training stability and avoid too large policy updates, PPO-Clip \cite{DBLP:journals/corr/SchulmanWDRK17} can be applied in MAPPO, and the surrogate objective for agent $i$ is defined as:
\begin{equation}
    J^{CLIP}(\theta^i)=E_{\bm{s}_t,\bm{a}_t}[min(r_t^i(\theta^i)\hat{A}(\bm{s}_t,\bm{a}_t),g(\epsilon,\hat{A}(\bm{s}_t,\bm{a}_t)))]
\end{equation}
where
\begin{equation}
    g(\epsilon,\hat{A}(\bm{s}_t,\bm{a}_t))=
    \begin{cases}
        (1+\epsilon)\hat{A}(\bm{s}_t,\bm{a}_t), \hat{A}(\bm{s}_t,\bm{a}_t) \geq 0 \\
        (1-\epsilon)\hat{A}(\bm{s}_t,\bm{a}_t), \hat{A}(\bm{s}_t,\bm{a}_t) < 0
    \end{cases}
\end{equation}
$\epsilon$ is a hyperparameter that indicates how far the new policy can diverge from the previous policy, usually as 0.2. The ratio $r_t^i(\theta^i)$ is defined as 
\begin{equation}
    r_t^i(\theta^i)=\frac{\pi_{\theta^i}(a^i_t|s^i_t)}{\pi_{\theta^i_{old}}(a^i_t|s^i_t)}
\end{equation}
$\pi_{\theta^i}(a^i_t|s^i_t)$  and $\pi_{\theta^i_{old}}(a^i_t|s^i_t)$ are the new policy and old policy, respectively. The PPO-clip controls the $r_t^i(\theta^i)$ within the interval $[1-\epsilon,1+\epsilon]$ to prevent the new policy from changing greatly compared to the old policy.

For a set of trajectories $D=\{\tau\}$, the parameter $\theta^i$ of the policy network of agent $i$ is updated by maximizing the PPO-Clip objective:
\begin{equation}
    \theta^i = \argmax_{\theta^i}\frac{|D|T}{1}\sum_{\tau \in D}\sum_{t=0}^{T-1}J^{CLIP}(\theta^i)
    \label{eq21}
\end{equation}

The parameter $\phi^i$ of the value network is updated by minimizing the mean-squared error:
\begin{equation}
\begin{split}
    \phi^i &= \argmin_{\phi^i}L(\phi^i) \\
    &=\argmin_{\phi^i}\frac{1}{|D|T}\sum_{\tau \in D}\sum_{t=0}^{T-1}(V_{\phi^i}(\bm{s_t})-\hat G_t)^2
    \label{eq22}
\end{split}
\end{equation}
The $\theta^i$ and $\phi^i$ are updated via stochastic gradient ascent and descent, respectively: 
\begin{equation}
    \theta^i \leftarrow \theta^i +\alpha \nabla_{\theta^i}J^{CLIP}(\theta^i)
\end{equation}
\begin{equation}
    \phi^i=\phi^i-\beta \nabla_{\theta^i}L(\phi^i)
\end{equation}
where $\alpha$ and $\beta$ are the learning rates of the policy network and value network, respectively.
\section{Viewpoint Prediction Performance Evaluation}
\subsection{Implementation Details}
\begin{figure}[htbp]
\centering
\includegraphics[width=0.4\textwidth]{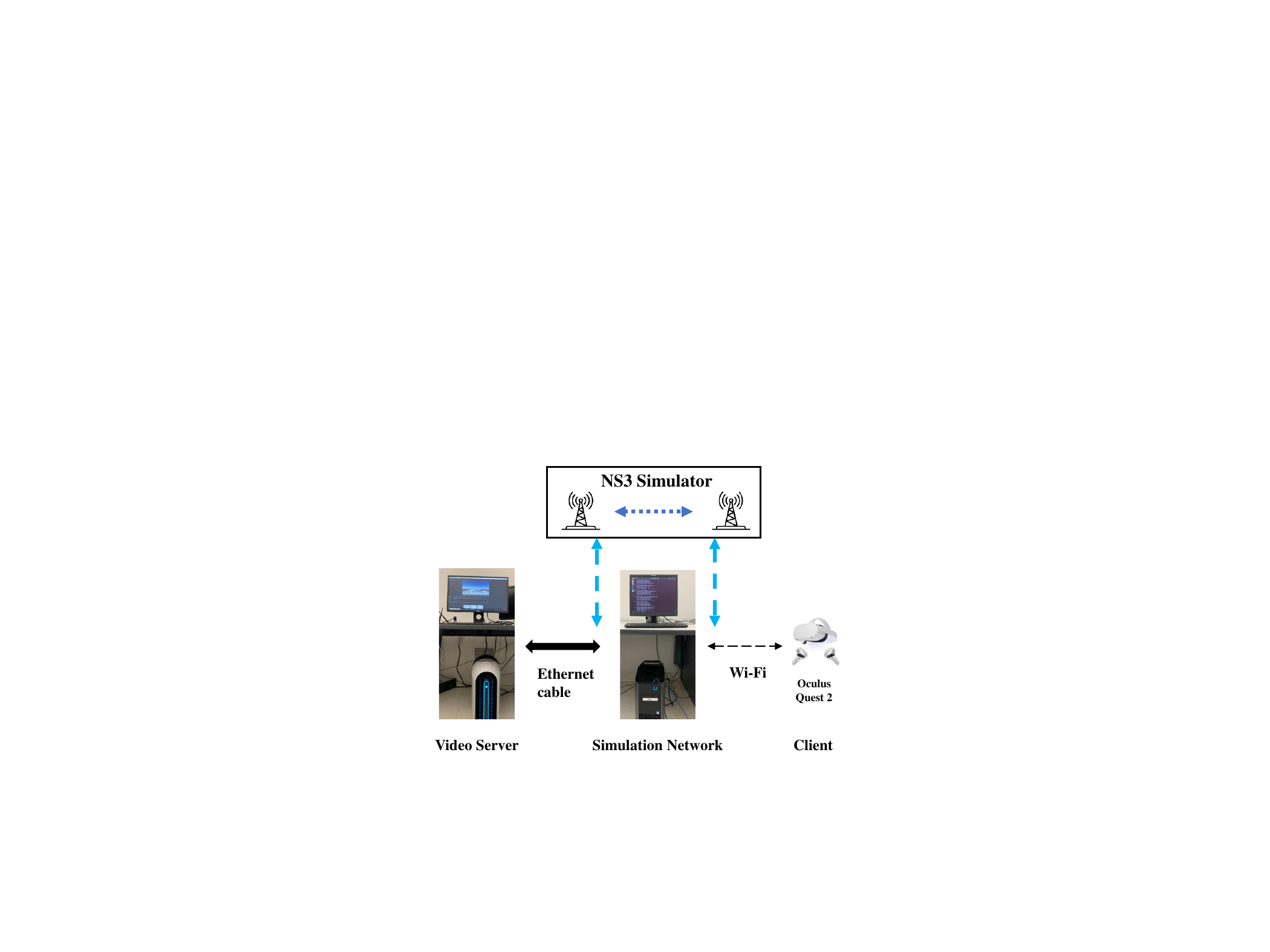}
\caption{The testbed for our proposed system contains a 360$\degree$ video server, a simulated network, and a client.}
\label{testbed}
\end{figure}

System setup: To verify our system, a testbed for our proposed ABR streaming system is built, as shown in Fig. \ref{testbed}. It mainly consists of a 360$\degree$ video server, a client, and a simulated network using NS3 \cite{riley2010ns}. We select one widely used 360$\degree$ video dataset, MMSys18 \cite{10.1145/3204949.3208139}, for a fair evaluation. MMSys18 contains the viewpoint trajectories of 57 participants watching 19 360$\degree$ videos. The 360$\degree$ videos are split into segments with a 1-second duration, and each segment is divided into $6 \times 12 =72$ tiles. The MADRL has three agents for three future viewpoint trajectories in different tiles.

\begin{figure}[htbp]
\centering
\includegraphics[width=0.45\textwidth]{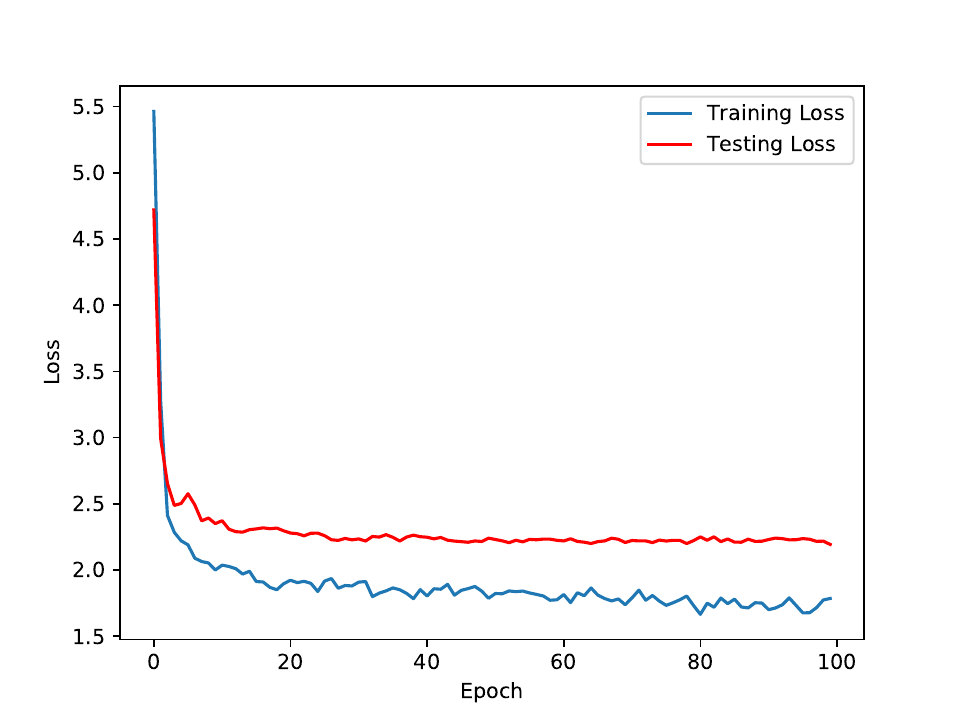}
\caption{The training and testing loss curves of the proposed transformer.}
\label{curve}
\end{figure}

Transformer model training: All experiments are conducted using a testbed built on a computer with an Intel Core i9-11900F CPU and an Nvidia GTX3090 GPU.
The time window of the input is set to 1 second. The model is tested with different output time windows from 1 to 5 seconds. Nonetheless, our system uses the 1-second future output. Five points and frames are sampled every second. Therefore, the model input has a sequence of 5 points and a sequence of 5 frames. The coordinates of viewpoints are normalized to the unit sphere and classified into 1500 clusters. Each viewpoint is represented by its cluster's centroid. Therefore, the input of the viewpoint encoder is a 1-second viewpoint trajectory containing 5 centroids instead of the original viewpoints. The 1500 clusters are used as labels with one-hot encoding for the transformer model. The visual encoder has 4 temporal and spatial encoder blocks while the viewpoint encoder has 2 transformer encoder blocks. The viewpoint decoder has 2 decoder blocks. The model uses 12 multi-head attention and an embedding dimension of 768. The cross-entropy loss function is used. The transformer is trained in 100 epochs. The batch size is 100, the learning rate is 0.0005 with a decay rate of 0.99, and the Adam optimization algorithm is used. The training and testing loss curves shown in Fig. \ref{curve} indicate our model fits well.  

\subsection{Results and Analysis for Viewpoint Prediction}
In our experiments, the great circle distance \cite{10.1145/3524273.3528176} and average great circle distance \cite{marchetti2020mantra} are used as the evaluation metrics. The great circle distance is the shortest distance at a specific time step between the ground truth point and the predicted point on the surface of a sphere, while the average great circle distance indicates the average error over all future timesteps. Since our method predicts multiple trajectories, we adopt the BMS (best of many samples) strategy \cite{10.1145/3524273.3528176} to evaluate our method, where the minimum error between $I$ predicted trajectories and ground truth is used as our result. The proposed method is compared to five existing methods. The baseline method uses the last input element as the output trajectory. The VPT360 \cite{9733647} uses the encoder of the transformer model without the decoder to predict the viewport trajectory. The Track \cite{9395242} is based on the LSTM. The DVMS \cite{10.1145/3524273.3528176} is proposed to predict multiple-viewpoint trajectories based on the discrete variational multiple sequence method. MANTRA \cite{10.1145/3524273.3528176} is proposed to predict multiple trajectories for vehicles using memory-augmented networks. Here, we modify the MANTRA to predict multiple-viewpoint trajectories. The notation *-1, *-2, and *-3 indicate the number of predicted trajectories for method *. For example, Ours-1, Ours-2 and Ours-3 indicate that our proposed method predicts 1, 2 and 3 trajectories, respectively. 

\begin{figure}[htbp]
\centering
\includegraphics[width=0.47\textwidth]{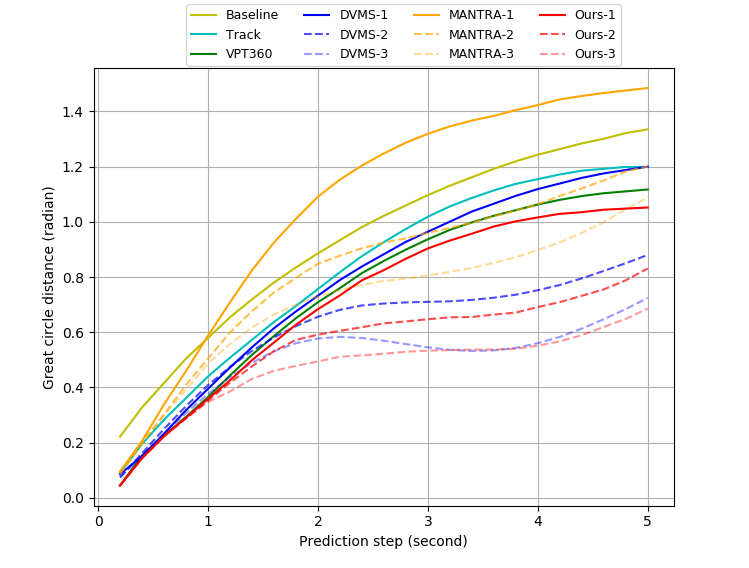}
\caption{ Great circle distance for different methods on dataset MMSys18. The notation *-1, *-2, and *-3 indicate the number of predicted trajectories for method *.}
\label{mmsys}
\end{figure}
As shown in Fig. \ref{mmsys}, when the number of trajectories $I$ is 1, our method achieves the best result. As the number of trajectories increases, the great circle distance decreases for DVMS, MANTRA, and the proposed method. When the number of trajectories is 3, the proposed method outperforms the existing methods. Although the distance of our method is similar to that of the DVMS during 3-4 seconds when the number of trajectories $I$ is 3, the proposed method performs better most of the time.   

\begin{table}[h!]
    \caption{Comparative analysis of viewpoint prediction based on average great circle distance.}
    \centering
    \begin{tabular}{ |c| c| c|c|c|c| }
    \hline
\multirow{2}{*}{Method}                        & \multicolumn{5}{c|}{Prediction Time Window}                                                                                                                                                                                                                                \\ \cline{2-6} 
                                               & \multicolumn{1}{c|}{1st s}                              & \multicolumn{1}{c|}{2nd s}                              & \multicolumn{1}{c|}{3rd s}                              & \multicolumn{1}{c|}{4th s}                              & 5th s                              \\ \hline
     Baseline & 0.409 & 0.592 & 0.734 & 0.848 & 0.939\\  
     \hline
     VPT360 \cite{9733647} & 0.215 & 0.399 & 0.550 & 0.667 & 0.754 \\
     \hline
     Track \cite{9395242} & 0.357 & 0.526 & 0.674 & 0.789 & 0.870 \\
     \hline
    DVMS-1 \cite{10.1145/3524273.3528176} & 0.246 & 0.432 & 0.583 & 0.703 & 0.796 \\
     \hline
     DVMS-2 & 0.245 & 0.409 & 0.506 & 0.562 & 0.614 \\
     \hline
      DVMS-3 & 0.228 & 0.373 & \underline{0.438} & \underline{0.464} & \underline{0.501} \\
     \hline
     MANTRA-1 \cite{10.1145/3524273.3528176} & 0.337 & 0.624 & 0.830 & 0.969 & 1.068 \\
     \hline
          MANTRA-2  & 0.299 & 0.516 & 0.651 & 0.743 & 0.825 \\
     \hline
          MANTRA-3  & 0.292 & 0.473 & 0.576 & 0.646 & 0.717 \\
     \hline
     Ours-1 & 0.212 & 0.386 & 0.531 & 0.643 & 0.723 \\
     \hline
     Ours-2 & \underline{0.210} & \underline{0.364} & 0.452 & 0.505 & 0.557 \\
     \hline
          Ours-3 & \textbf{0.209} & \textbf{0.329} & \textbf{0.393} & \textbf{0.430} & \textbf{0.468} \\
     \hline
    \end{tabular}
    \label{tf_results}
\end{table}

The experimental results on the average great circle distance are shown in Table \ref{tf_results}. The bold numbers are the best scores, and underlined are the second-best scores. The proposed method has the smallest average error across all time windows when three trajectories are predicted.

\begin{figure}[htbp]
\centering
\includegraphics[width=0.43\textwidth]{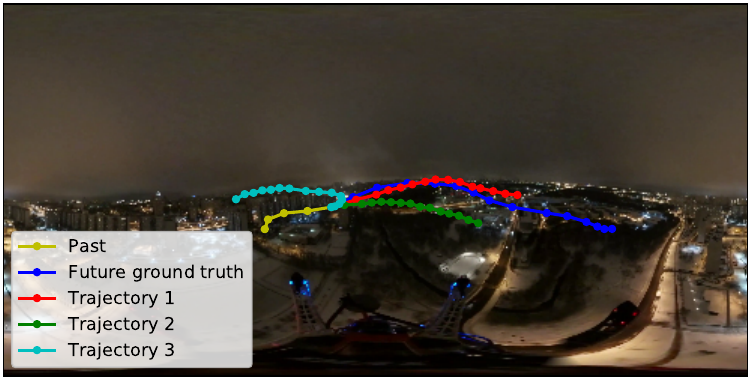}
\caption{An example of predicted multiple trajectories by the transformer.}
\label{multi}
\end{figure}

Our proposed method can predict multiple future trajectories with their probabilities. An example of multiple viewpoint trajectories predicted by our method is shown in Fig. \ref{multi}. When the 1-second past viewpoint trajectory is fed into the transformer, three possible future trajectories are obtained with their viewing probabilities. As shown in Fig. \ref{multi}, trajectory 1 matches the ground truth very well, while others cover the user's other possible trajectories, which are learned by the transformer from the viewpoint trajectories of other users.

\subsection{Ablation Study}
\subsubsection{The Multi-Viewpoint Prediction Necessity Analysis}
The work of DVMS \cite{10.1145/3524273.3528176} analyzes the necessity of multiple predictions from two perspectives: data and model. They find that relatively close past trajectories may lead to distinct/farther apart future trajectories. The single-viewpoint prediction model is less accurate when there is diversity in the true futures. This diversity of futures indeed contributes to the error of the single-viewpoint model.
\begin{table}[h!]
    \caption{Comparative analysis of viewpoint prediction based on average great circle distance for sudden movement.}
    \centering
    \begin{tabular}{ |c| c| c|c|c|c| }
    \hline
\multirow{2}{*}{Method}                        & \multicolumn{5}{c|}{Prediction Time Window}                                                                                                                                                                                                                                                                                                    \\ \cline{2-6} 
                                               & \multicolumn{1}{c|}{1st s}                              & \multicolumn{1}{c|}{2nd s}                              & \multicolumn{1}{c|}{3rd s}                              & \multicolumn{1}{c|}{4th s}                              & 5th s                              \\ \hline
     Baseline & 0.832 & 0.911 & 1.083 & 1.187 & 1.275\\  
     \hline
     VPT360 \cite{9733647} & 0.548 & 0.668 & 0.732 & 0.875 & 0.954 \\
     \hline
     Track \cite{9395242} & 0.693 & 0.764 & 0.876 & 0.957 & 1.015 \\
     \hline
    DVMS-1 \cite{10.1145/3524273.3528176} & 0.557 & 0.693 & 0.764 & 0.847 & 0.918 \\
     \hline
     DVMS-2 & 0.468 & 0.629 & 0.693 & 0.750 & 0.802 \\
     \hline 
      DVMS-3 & \underline{0.418} & \underline{0.559} & \underline{0.598} & \underline{0.674} & \underline{0.746} \\
     \hline
     MANTRA-1 \cite{10.1145/3524273.3528176} & 0.754 & 0.848 & 0.981 & 1.119 & 1.241 \\
     \hline
          MANTRA-2  & 0.661 & 0.757 & 0.849 & 0.925 & 1.153 \\
     \hline
          MANTRA-3  & 0.605 & 0.649 & 0.757 & 0.760 & 0.772 \\
     \hline
     Ours-1 &0.525 & 0.637 & 0.794 & 0.839 & 0.893 \\
     \hline
     Ours-2 & 0.473 & 0.592 & 0.650 & 0.691 & 0.765 \\
     \hline
          Ours-3 & \textbf{0.412} & \textbf{0.551} & \textbf{0.594} & \textbf{0.645} & \textbf{0.721} \\
     \hline
    \end{tabular}
    \label{sudden}
\end{table}

We analyze the model's ability to handle sudden movement data. We choose sudden movement trajectories from the testing set of MMSys18. At time step $t$, if the speed change between the current and subsequent time steps exceeds 5 times, a 6-second trajectory from 1 second before to 5 seconds after $t$ is selected. As shown in Table \ref{sudden}, the prediction error on sudden movement trajectories becomes larger compared to results on the whole dataset shown in Table \ref{tf_results}. However, the multi-viewpoint prediction significantly reduces the distance compared to single-viewpoint prediction, which indicates the multi-viewpoint prediction has a strong ability to handle sudden movement data.

\subsubsection{Generalizability Analysis}
The proposed transformer uses attention mechanisms and data fusion to capture the spatial and temporal information of the video content and viewpoint trajectories, which allows the model to focus on relevant parts of the video and trajectory across space and time dynamically and improve the model’s generalizability. Additionally, the model is trained and evaluated in a challenging dataset MMSys18 containing the viewpoint trajectories of 57 participants watching 19 different types of {360\degree} videos, which enables the model’s strong generalizability.
Furthermore, data augmentation and regularization techniques are also used during training to enhance the model's generalizability, such as frame resizing for videos, and position rescaling for trajectories.

\subsubsection{Limitation Analysis}
The proposed transformer model captures spatial-temporal information from video frames and viewpoint trajectories to enhance prediction performance. However, this enhancement comes at the expense of considerable computational resources and time. The model exhibits a runtime of approximately 63.1 milliseconds for predicting a 1-second window, indicating the runtime is acceptable.

{\section{Rate Adaptation Performance Evaluation}
\subsection{Implementation Details}
Network traces: Two public network trace datasets, FCC \cite{fcc} and HSDPA \cite{10.1145/2483977.2483991}, are used to evaluate our proposed method and other ABR methods in real network conditions. A total of 36,020 traces are available, with 80\% and 20\% of the traces randomly selected as the training and test sets, respectively. As few network traces can not support the lowest bitrate, the network throughput is enlarged by 3 Mbps.

Viewpoint trajectory: Our method is based on multi-viewpoint prediction, and there is no public multi-viewpoint trajectory dataset. To evaluate our method, we use the single-viewpoint trajectory from the MMSys18 testing set as input to the transformer for the generation of multiple viewpoint trajectories. The MADRL model assigns a bitrate to each viewport based on the predicted trajectories of the transformer. The following results are obtained from the proposed transformer and MADRL method.

QoE objectives: Since different parameters in QoE represent different preferences, four sets of weighting parameters are selected for $(\alpha_1, \alpha_2, \alpha_3, \alpha_4)$ to represent four different QoE objectives, where (1,1,1,1), as the baseline, indicates maximizing the quality of the viewport regions as the viewport quality contributes more to QoE. As the viewport quality has a direct impact on the user's experience, without loss of generality, we weigh other factors 2 to keep high quality and reduce the desired factors simultaneously. Therefore, QoE objectives (1,2,1,1), (1,1,2,1), and (1,1,1,2) represent emphasizing minimizing the temporal quality variation, spatial quality variation, and rebuffering time, respectively. 

MADRL model training: The training of the MADRL model is conducted on the same computer used for the transformer. For actors and critics, the observation is concatenated to an array fed into a CNN. The learning rate is 0.0001. The discount factor $\gamma$ is 0.99, and the hyperparameter $\lambda$ is 0.95.

\subsection{Methods for Comparison}
We compare our proposed strategy with several state-of-the-art strategies, as shown in the following:
\begin{itemize}
    \item Buffer-based (BB) \cite{10.1145/2619239.2626296}: This method selects bitrate for the entire 360$\degree$ video using the buffer-based algorithm, which determines bitrates to maintain the buffer occupancy at least 5 seconds. If the buffer occupancy exceeds 15 seconds, it will automatically select the highest available bitrate.
    \item MPC \cite{10.1145/2785956.2787486}: This method determines the same bitrate for all viewport regions by considering the buffer level information and throughput prediction to maximize the given QoE.
    \item Dynamic \cite{pengviewport}: This method determines the same bitrate for all viewports. If the buffer level is above the threshold, BOLA \cite{9110784} is used to make bitrate decisions; otherwise, RB \cite{10.1145/1943552.1943575} is used.
    \item Pensieve \cite{10.1145/3098822.3098843}: This method is proposed for traditional video streaming. A SADRL model with an A3C algorithm is employed to determine bitrates. Here, we use this method to determine the same bitrate for all viewport regions of the 360$\degree$ video using a SADRL model with A3C.
    \item RAPT360 \cite{9419061}: This method utilizes A3C to determine the bitrate for all tiles simultaneously, which is used here to determine the bitrate for viewport regions.
    \item Feng$\_$PPO \cite{9838819}: This method uses PPO to determine the bitrate for the tiles inside the viewport. Here, we use it to determine bitrate for viewports simultaneously.
    \item DRL360 \cite{8737361}: This method uses a SADRL with A3C to select the same bitrate for tiles within the viewport while assigning the lowest bitrate to other tiles outside the viewport. We use this method to select the bitrate for multiple regions one by one.
    \item 360SRL \cite{9234071}: Instead of determining the bitrate of all tiles at once, this method selects the bitrate for each tile in sequence by using SADRL with A3C.
    \item Independent PPO (IPPO) \cite{DBLP:journals/corr/abs-2011-09533}: The IPPO is a decentralized MADRL algorithm, where the agents of the IPPO do not share information. Each agent receives the global reward and updates policies based on its local observation. 
\end{itemize}

\subsection{Quality-of-Experience Analysis}

The experimental results in Fig. \ref{result} show that the proposed method outperforms other existing ABR methods under various QoE objectives. specifically, the proposed method is superior to existing algorithms when users prefer high video quality, small temporal variations, small spatial variations, and small rebuffed time, as shown in Fig. \ref{result} a, b, c, and d, respectively. The QoE of the proposed method is improved by about 4.8\%-85.5\%, 3.6\%-81.5\%, 1.7\%-83.9\%, and 4.5\%-84.4\% for QoE objective (1,1,1,1), (1,2,1,1), (1,1,2,1) and (1,1,1,2), respectively. Particularly, for QoE objective (1,1,1,1), the proposed method outperforms BB, MPC, Dynamic, Pensieve, RAPT360, Feng$\_$PPO, 360SRL, DRL360, and IPPO by 85.5\%, 61.9\%, 67.6\%, 57.7\%, 12.9\%, 11.1\%, 7.5\%, 8.0\%, and 4.8\% in terms of the normalized average QoE, respectively.

\begin{figure}[ht]
\centering
\includegraphics[width=0.49\textwidth]{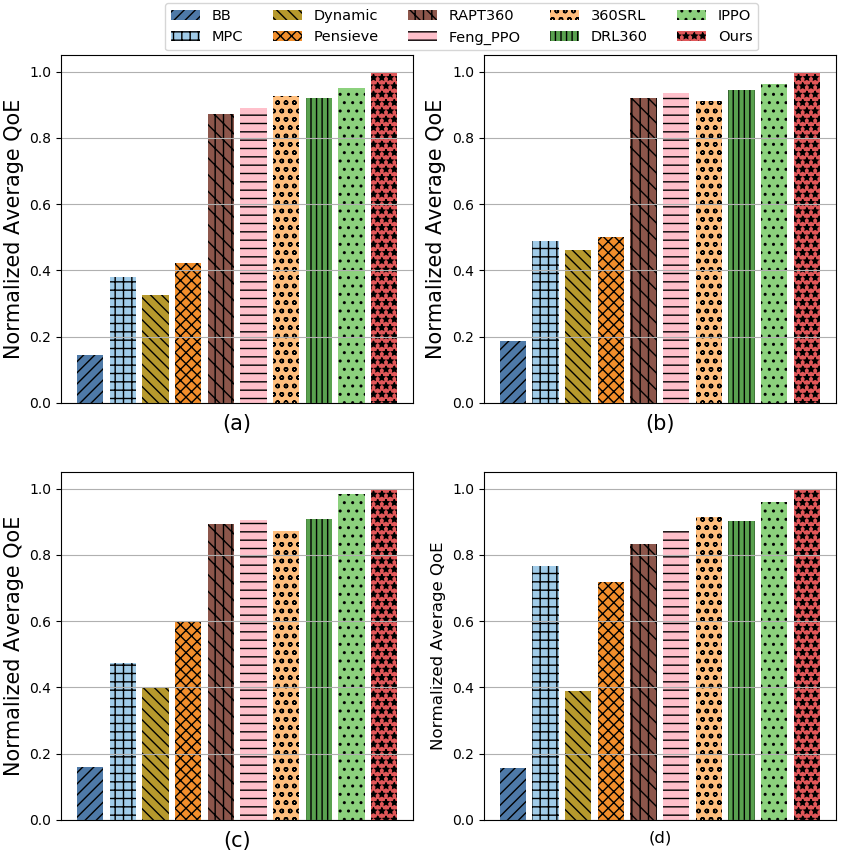}
\caption{Comparing our method with existing ABR algorithms over four different QoE objectives: (a) (1,1,1,1), (b) (1,2,1,1), (c) (1,1,2,1), (d) (1,1,1,2).}
\label{result}
\end{figure}

\begin{figure}[h]
\centering
\includegraphics[width=0.49\textwidth]{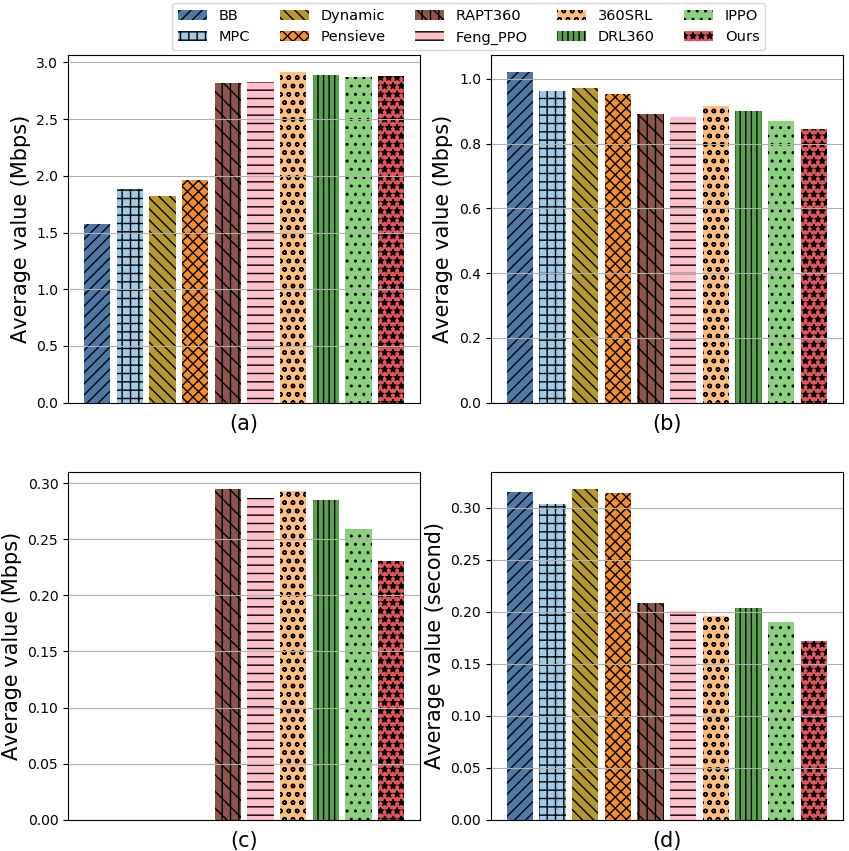}
\caption{Comparing our method with existing ABR algorithms in terms of (a) viewport quality, (b) viewport temporal variation, (c) viewport spatial variation, and (d) rebuffering, with the QoE objective (1,1,1,1). In (c), since BB, MPC, Dynamic, and Pensieve adopt the same bitrate for all viewport regions, they have zero spatial quality variation.} 
\label{qoe breakdown}
\end{figure}

\begin{figure*}[ht]
\centering
\includegraphics[width=1.0\textwidth]{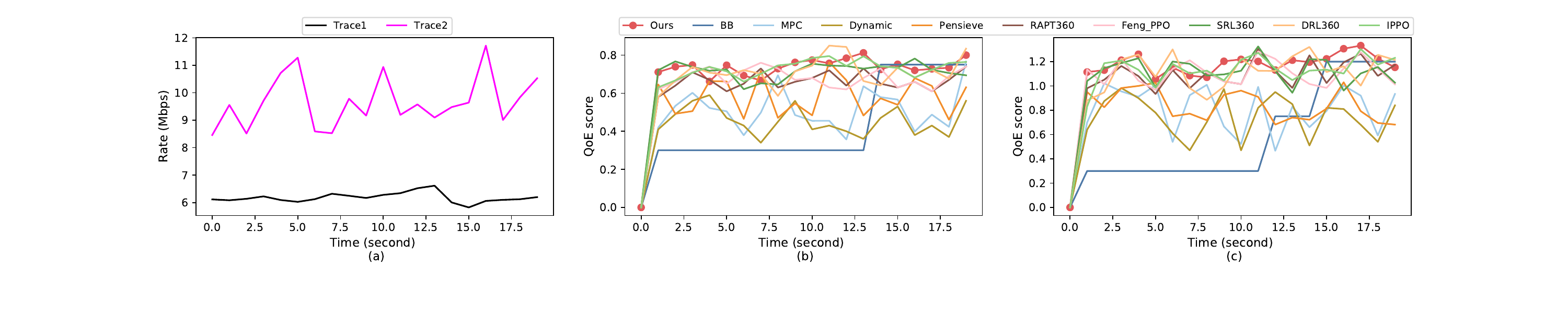}
\caption{The QoE visualization of the proposed method on two network traces with the QoE objective (1,1,1,1): (a) Network traces, (b) QoE score on trace 1, (c) QoE score on trace 2.}
\label{vis}
\end{figure*}
For all QoE subjects, the BB method determines bitrate by only considering the buffer occupancy without the QoE metric. Therefore, it keeps the same QoE score under various QoE objectives. Meanwhile, the BB method has the worst performance compared to other methods, as it is used to determine the bitrate for the entire 360$\degree$ video without considering the viewport and tile, which leads to a lot of bandwidth resources being wasted in the tiles outside the viewport regions. The MPC, Dynamic, and Pensieve determine the same bitrate for all viewport regions, ignoring differences in the probability that users view various regions, which also causes the bandwidth resource not to be well utilized in the regions with low viewing probability. The QoE scores of RAPT360, Feng$\_$PPO, SRL360, DRL360, IPPO, and the proposed method are much better compared to BB, MPC, Dynamic, and Pensieve under different QoE objectives, which shows the advantage of adaptive bitrate determination for each tile considering viewport probability. Additionally, the performance of IPPO is close to the proposed method and outperforms other SADRL-based methods for most QoE objectives, indicating the advantage of the MADRL. IPPO's performance is not as good as ours since all agents are independent and do not communicate with each other. We also notice that the improvement of the MAPPO is not much compared to the IPPO. The reason may be the complex network environment because the methods are evaluated on two real-world network datasets with a total of 36020 traces. Additionally, the only partially observable nature of these environments poses a challenge for both MAPPO and IPPO in learning effective coordination strategies. Although the CTDE utilizes global information to improve MAPPO's performance during training, the performance during execution is still constrained by their limited observation, resulting in limited advancements over IPPO.

To figure out the average QoE gains obtained by the proposed method, the performance of each method on the individual terms of the definition of the QoE  (Eq. \ref{qoe}) is analyzed. Specifically, in the QoE objective (1,1,1,1), the proposed method is compared to the ABR algorithms, as shown in Fig. \ref{qoe breakdown}, in terms of the average viewport quality, the penalty from temporal quality variation, the penalty from spatial quality variation, and the penalty from rebuffering. Since BB, MPC, Dynamic, and Pensieve adopt the same bitrate for all viewport regions, they do not have a spatial quality variation between viewport regions. However, they have lower video quality than other methods because of insufficient bandwidth utilization. Moreover, all of them suffer higher temporal quality variation penalties and rebuffering penalties. For example, the temporal quality variation penalty of the proposed method is 0.84 Mbps improved by 17.6\%, 12.5\%, 13.4\%, 11.6\%, 6.0\%, 4.5\%, 8.7\%, 6.7\%, 3.6\%, respectively compared to BB (1.02 Mbps), MPC (0.96 Mbps), Dynamic (0.97 Mbps), Pensieve (0.95 Mbps), RAPT360 (0.89 Mbps), Feng$\_$PPO (0.88 Mbps), 360SRL (0.92 Mbps), DRL360 (0.90 Mbps), and IPPO (0.87 Mpbs).

SADRL-based methods only optimize the QoE of the current viewport by maximizing the video quality while ignoring the existence of the other viewport regions, which causes higher penalties on temporal variation, spatial variation, and rebuffering. Compared to other methods, the performance gains of the proposed method are largely attributed to its ability to limit rebuffering across various networks. As a result, rebuffering of the proposed method is reduced by 45.4\%, 43.2\%, 45.9\%, 45.2\%, 17.3\%, 14.4\%, 12.2\%, 15.3\%, and 9.5\% compared to BB, MPC, Dynamic, Pensieve, RAPT360, Feng\_PPO, 360SRL, DRL360, and IPPO, which ensures that a sufficient buffer is built up to handle fluctuations in throughput that occur in the network. It can be seen that the proposed method is able to balance each factor in a way that optimizes the QoE metric instead of performing the best on every QoE element compared to other ABR methods. For instance, compared with 360SRL, the proposed method has about 1.3\% lower in video quality but 8.7\% higher on the temporal variation penalty. 

\subsection{QoE Visualization on Network Traces}
Furthermore, an evaluation of the adaptability of QoE to bandwidth over time is conducted for a more comprehensive comparison of our method with other methods, as shown in Fig. \ref{vis}. We measure the QoE score over two typical network conditions. As shown in Fig. \ref{vis} (a), trace 1 has characteristics with a lower bit rate and less fluctuation, where the mean is about 4.96 Mbps, and STD (standard deviation) is about 1.17 Mbps. While trace 2 has characteristics of a higher bit rate and more fluctuation, where the mean is about 6.91 Mbps and STD is about 3.37 Mbps. The QoE score on the trace 1 is shown in Fig. \ref{vis} (b). Since the BB only considers the buffer occupancy and streams 360$\degree$ video as a traditional video to determine bitrate, it presents stability during a period. Meanwhile, it can be seen that the MPC, Dynamic, and Pensieve have very large fluctuations for both traces, as they need to change the bitrate frequently for all viewport regions by considering the viewport probabilities to meet various network conditions. For trace 1, the average QoE score of the proposed method is about 0.74 improved by around 40.2\%, 30.5\%, 38.5\%, 21.0\%, 10.2\%, 7.9\%, 2.8\%, 3.8\%, 1.4\% compared with BB, MPC, Dynamic, Pensieve, RAPT360, Feng\_PPO, 360SRL, DRL360, and IPPO, respectively. The STD of the QoE score for our method is about 0.04, which decreases by about 13.6\%-81\%. For the QoE score on trace 2, as shown in Fig. \ref{vis} (c), all methods present higher QoE scores with higher fluctuation compared to the results of trace 1 due to the characteristics of trace 2. The mean of the QoE score of the proposed method on trace 2 is about 1.18, which is improved by about 4.4\%-93.4\% compared to other methods. The STD of the proposed method is about 0.07, which increases by 23.6\%. The fluctuation of the QoE score is mainly caused by bandwidth fluctuations and viewpoint prediction probabilities. The results of the trace 2 are better than that of the trace 1, as expected from the characteristic difference between the two traces. Although the proposed method does not have the best performance over time under different network conditions, it can stably provide a high average QoE value.

\subsection{Evaluation by Other Metrics}
Besides the QoE and its breakdown metrics (viewport quality, quality temporal variation, quality spatial variation, and rebuffering time), there are some other assessment metrics worth measuring and discussing. Thus, we further evaluate the system performance using three different metrics: latency \cite{9069299,8865443}, playback freeze frequency \cite{9833344,10.1145/3394171.3413751}, and training and testing time. The results for the three metrics with different methods are shown in Table \ref{frequency}.

\begin{table}[htbp]
\centering
\caption{Latency, playback freeze frequency and training and testing time for different methods.}
\label{frequency}
\resizebox{0.4\textwidth}{!}{%
\begin{tabular}{|c|c|c|c|c|}
\hline
Method       & \begin{tabular}[c]{@{}c@{}}Latency \\ (ms)\end{tabular}    & \begin{tabular}[c]{@{}c@{}}Playback freeze  \\ frequency\end{tabular}   &  \begin{tabular}[c]{@{}c@{}}Training \\  time (hr)  \end{tabular}  & \begin{tabular}[c]{@{}c@{}}Testing \\   time (min)  \end{tabular}   \\ \hline
BB           &         114       & 0.208  & ---   & ---   \\ \hline
MPC          &        103                 & 0.197    &  ---  & ---     \\ \hline
Pensieve        &   96                        & 0.184   & 0.7 &9.8          \\ \hline
Dynamic           &    106                   & 0.202    & ---  & ---       \\ \hline
RAPT360  &    81 & 0.154       &2.1 &11.2    \\ \hline 
Feng\_PPO &    79    & 0.154    & 2.3 &11.2  \\ \hline 
360SRL           &  81           & 0.153   &  1.6 &33.4               \\ \hline
DRL360            &   80                     & 0.151   &0.9 & 28.9                    \\ \hline
IPPO   &    77   & 0.149  &1.3   &24.5  \\ \hline
Ours             &    75         & 0.148     &1.2 &24.5          \\ \hline

\end{tabular}%
}
\end{table}
The latency shows the end-to-end delay from the server to the client, which directly impacts the user's experience. The latency of our proposed method is 75 milliseconds, which outperforms others by 3\%-34\%. The DRL-based methods have lower latency compared to traditional rule-based methods. The proposed approach can offer users a higher quality of experience with low latency, compared with other approaches, because it allocates lower bitrates to regions of low importance and higher bitrates to regions of high importance.

Playback freeze frequency indicates the number of freezes during the 360$\degree$ video playback. High-frequency playback freeze results in a worse user experience. As shown in Table \ref{frequency}, the proposed method has the lowest freeze frequency (0.148), followed by the IPPO (0.149), which demonstrates that our method with global information makes better decisions for 360$\degree$ video transmission. The proposed method reduces the number of freezes by 28.8\% compared to BB. The DRL-based methods perform better than traditional rule-based methods, which shows DRL-based methods have more flexibility to adjust the video bitrate according to various network conditions.

The training time is the duration needed for the model to converge, while the testing time indicates the duration required to evaluate the performance of the model on the test dataset. The training and testing time provide insights into the efficiency and scalability of the model. The training time and testing time of our proposed method are 1.2 hours and 24.5 minutes, respectively. The Pensieve achieves the lowest QoE among DRL-based methods despite taking the shortest training and testing time. The DRL360 takes 0.9 hours to train but requires longer testing time compared to our method. Similarly, the RAPT360 and Feng$\_$PPO have shorter testing times but longer training times compared to ours. By jointly considering the training and testing time, our method demonstrates good performance. In addition, our model has an inference time of approximately 10 milliseconds, satisfying the real-time demands of practical applications.}

\subsection{Ablation Study:}
\subsubsection{Impact of Hyperparameters}
\begin{figure}[H]
\centering
\includegraphics[width=0.45\textwidth]{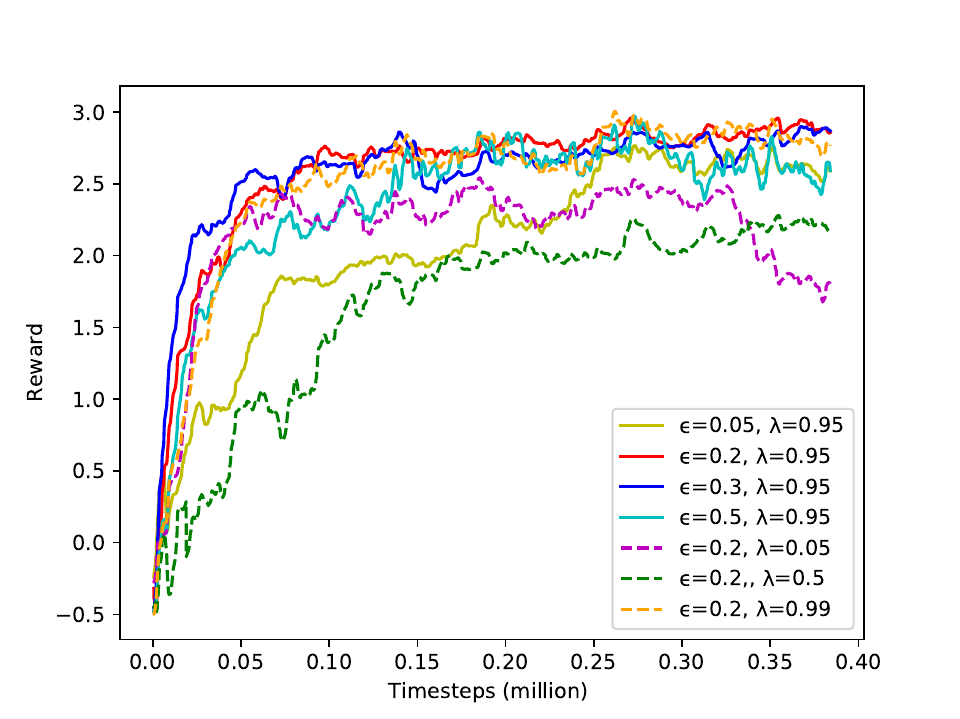}
\caption{The impact of different values for hyperparameters $\epsilon$ and $\lambda$ on the reward of QoE objective (1,1,1,1).}
\label{hyper}
\end{figure}

Due to the importance of hyperparameters in model performance, we investigate the impact of two hyperparameters, $\epsilon$ and $\lambda$, on QoE. By setting different values for the $\epsilon \in \{0.05, 0.2, 0.3, 0.5\} $ and $\lambda \in \{0.05, 0.5, 0.95, 0.99\}$, different QoE rewards are obtained and shown in Fig. \ref{hyper}. The proposed method achieves the best performance when $\epsilon =$ 0.2 and $\lambda = 0.95$, which are the values used in our work. The settings of ($\epsilon =$0.3,  $\lambda=$0.95) and ($\epsilon =$0.2,  $\lambda=$099) also converge quickly, but with lower rewards. Other settings are characterized by low rewards, slow convergence, and even overfitting.

\subsubsection{Impact of Network Structure}
Since MADRL has two neural networks, the structure of the network has a very important influence on the performance of the proposed method. Hence, we modify the number of convolution layers in the CNN between 1 and 6 to examine the effect of the network structure of the proposed method on the average quality of experience. As shown in Table \ref{t2}, the network structure is evaluated from one layer with 16 channels to 6 layers with the channel configuration (16,32,64,128,256,512) for each layer. It can be seen that the proposed method with one layer structure has the lowest QoE score, as it is hard to deal with the observation information. As the number of CNN layers increases, the performance of the proposed method increases.

\begin{table}[htbp]
\centering
\caption{The impact of the number of CNN layers on the average QoE with configuration (1,1,1,1).}
\label{t2}
\scalebox{1}{
\begin{tabular}{|c|c|}
\hline
Number of Layer               & Average QoE \\ \hline
1 (16)               & 2.29           \\ \hline
2 (16,32)            & 2.47           \\ \hline
3 (16,32,64)         & 2.56           \\ \hline
4 (16,32,64,128)     & 2.63           \\ \hline
5 (16,32,64,128,256) & 2.83           \\ \hline
6 (16,32,64,128,256,512) & 2.76           \\ \hline
\end{tabular}}
\end{table}

The average QoE score is improved by about 7.6\% and achieves the highest value when the CNN has 5 layers compared to the results of other structures. The performance decreases when the network structure has 6 layers. Thus, we set the number of CNN layers to 5 in our experiments. Furthermore, we can observe that the network structure significantly impacts the proposed method's performance. The average QoE value with a 5-layer structure is improved by 20.5\% compared to the 1-layer structure. 

\subsubsection{Generalizability  Analysis}
To enhance the generalizability of the proposed MADRL method, we train the MADRL model on a large amount of complex {360\degree} video streaming environments with various real-world network conditions. Additionally, the CTDE is used to enhance the generalizability of the proposed method. During training, a centralized critic has access to the global state information and observations from all agents. This additional information can help the agents learn more robust and coordinated policies that consider the global context, rather than relying solely on their partial observations. The decentralized execution ensures that the policies are practical and scalable, capable of being applied in real-world settings where each agent operates based on its own observations. Thus, the CTDE framework enhances the generalizability and scalability of the proposed method.

\subsubsection{Limitation Analysis}
Centralized training can become computationally expensive and difficult to scale as the number of agents increases. The complexity arises from the need to process and integrate information across all agents during training, leading to increased memory and computational requirements. Meanwhile, the centralized training process requires significant communication overhead to aggregate observations, actions, and rewards from all agents. This can be particularly challenging in distributed training setups or environments with limited communication bandwidth.

We also recognize the DRL model may perform in fluctuation and sometimes the performance cannot be guaranteed. As we focus on the introduction of a novel MADRL-based ABR strategy for 360° video streaming in this paper, future efforts could explore additional strategies to address this challenge. A promising approach might be for the streaming system to implement a hybrid-control method that merges DRL with a rule-based strategy. This is particularly valuable when there is a significant deviation between the decisions made by the DRL model and those proposed by the rule-based method. In such scenarios, leveraging the rule-based method's outputs can act as a counterbalance to stabilize the system and prevent large fluctuations.
\subsubsection{Security Concerns}
As our system collects and utilizes user data for multi-viewpoint prediction and rate adaptation, it is imperative to address privacy and security concerns. The system only gathers essential data listed in Fig. \ref{ctde} and avoids the collection of unnecessary personal or sensitive information. The viewpoint trajectory collected from the user is processed locally by the transformer on the client, as shown in Fig. \ref{view}, reducing transmission over the network.  
The system removes personally identifiable information from the collected data, ensuring that the data cannot be traced back to individual users.
During transmission between server and client, encryption technique (TLS) is used to protect against unauthorized access and data breaches.

\section{Conclusion}
In this paper, to handle the uncertainty of the user's head movement, we propose a multi-viewpoint trajectory prediction method, where a multimodal spatial-temporal attention transformer utilizes information of video frames and viewpoint trajectories, and treats the prediction problem as a classification to predict multiple future trajectories with their probabilities. Based on the multiple predictions, we propose a MADRL-based approach to dynamically optimize the QoE for 360$\degree$ video under various bandwidth conditions. The optimization problem is formulated as a Dec-POMDP problem. The MAPPO based on CTDE is presented to solve the problem. To reduce the number of agents in MADRL, each agent determines the bitrate for each viewport region. Experimental results show that the proposed method outperforms other existing ABR methods under the defined QoE metric, latency, and playback freeze frequency.

\small
\bibliographystyle{IEEEtran}
\bibliography{ref.bib}

\begin{IEEEbiography}[{\includegraphics[width=1in,height=1.25in,clip,keepaspectratio]{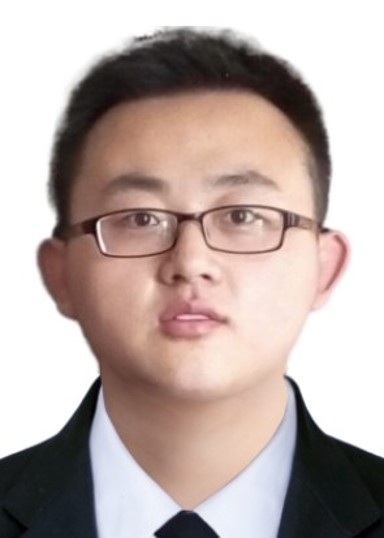}}]{Haopeng Wang}(hwang266@uottawa.ca) received the M.Eng. degree in electronic and communication engineering and B.Eng. degree in information and electronics from Beijing Institute of Technology, Beijing, China, in 2017 and 2015, respectively. He is currently pursuing the Ph.D. degree in electrical and computer engineering at the University of Ottawa. His research interests are computer network, extended reality, and multimedia.
\end{IEEEbiography}

\begin{IEEEbiography}[{\includegraphics[width=1in,height=1.25in,clip,keepaspectratio]{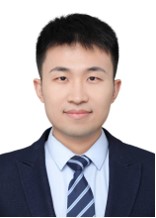}}]{Zijian Long}
(zlong038@uottawa.ca) received the B.Sc. degree in Software Engineering from Beijing Institute of Technology, China, in 2016 and the M.Sc. degree in Electrical and Computer Engineering from the University of Ottawa, Canada, in 2020. He is currently a Ph.D. candidate in the School of Electrical Engineering and Computer Science, University of Ottawa. His research interests include metaverse, XR network, and reinforcement learning.
\end{IEEEbiography}

\begin{IEEEbiography}[{\includegraphics[width=1in,height=1.25in,clip,keepaspectratio]{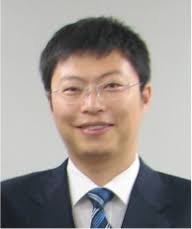}}]{Haiwei Dong}
(haiwei.dong@ieee.org) is currently a Director and Principal Researcher with Huawei Canada, and an Adjunct Professor with the University of Ottawa. He was a Principal Engineer with Artificial Intelligence Competency Center, Huawei Technologies Canada, Toronto, ON, Canada, a Research Scientist with the University of Ottawa, Ottawa, ON, Canada, a Postdoctoral Fellow with New York University, New York City, NY, USA, a Research Associate with the University of Toronto, Toronto, ON, Canada, and a Research Fellow (PD) with the Japan Society for the Promotion of Science, Tokyo, Japan. He received the Ph.D. degree from Kobe University, Kobe, Japan in 2010 and the M.Eng. degree from Shanghai Jiao Tong University, Shanghai, China, in 2008. He is a Senior Member of IEEE, a Senior Member of ACM, and a registered Professional Engineer in Ontario. His research interests include artificial intelligence, multimedia, metaverse, and robotics. He also serves as a Column Editor of IEEE Multimedia Magazine; an Associate Editor of ACM Transactions on Multimedia Computing, Communications, and Applications; and an Associate Editor of IEEE Consumer Electronics Magazine.
\end{IEEEbiography}

\begin{IEEEbiography}[{\includegraphics[width=1in,height=1.25in,clip,keepaspectratio]{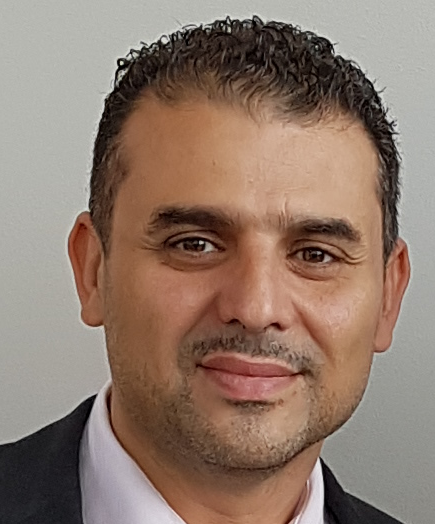}}]{Abdulmotaleb El Saddik}
(elsaddik@uottawa.ca) is currently a Distinguished Professor with the School of Electrical Engineering and Computer Science, University of Ottawa and a Professor with MBZUAI. He has supervised more than 120 researchers. He has coauthored ten books and more than 550 publications and chaired more than 50 conferences and workshops. His research interests include the establishment of digital twins to facilitate the well-being of citizens using AI, the IoT, AR/VR, and 5G to allow people to interact in real time with one another as well as with their smart digital representations. He received research grants and contracts totaling more than \$20 M. He is a Fellow of Royal Society of Canada, a Fellow of IEEE, an ACM Distinguished Scientist and a Fellow of the Engineering Institute of Canada and the Canadian Academy of Engineers. He received several international awards, such as the IEEE I\&M Technical Achievement Award, the IEEE Canada C.C. Gotlieb (Computer) Medal, and the A.G.L. McNaughton Gold Medal for important contributions to the field of computer engineering and science.
\end{IEEEbiography}
\end{document}